\documentclass[twocolumn,10pt]{asme2ej}

\usepackage{graphicx} 
\usepackage{hyperref}   
\usepackage{booktabs}
\usepackage{multirow}
\usepackage{array}
\usepackage{booktabs}
\usepackage{adjustbox}
\usepackage{ltablex}
\usepackage{caption, subcaption}
\usepackage{amsmath}
\usepackage{supertabular}
\usepackage[square,numbers]{natbib}

\usepackage{hyperref}

\hypersetup{
	colorlinks=true,
	linkcolor=blue,
	citecolor=blue,
	urlcolor=blue,
}

\title{DeepJEB: 3D Deep Learning-based Synthetic Jet Engine Bracket Dataset}

\author{Seongjun Hong\thanks{Equal contribution.} 
    \affiliation{
	AI Researcher\\ 
        Narnia Labs\\
        Daejeon 34051, South Korea\\
    Email: seongjun.hong@narnia.ai
    }	
}

\author{Yongmin Kwon$^*$
    \affiliation{
	AI Researcher\\
        Narnia Labs\\
        Daejeon 34051, South Korea\\
    Email: yongmin.kwon@narnia.ai
    }	
}

\author{Dongju Shin
    \affiliation{
	AI Researcher\\
        Narnia Labs\\
        Daejeon 34051, South Korea\\
    Email: dongju.shin@narnia.ai
    }	
}

\author{Jangseop Park
    \affiliation{
	PhD Student\\
	Cho Chun Shik Graduate School of Mobility\\
	Korea Advanced Institute of Science and Technology\\
        Daejeon 34051, South Korea\\
    Email: jangseop@kaist.ac.kr
    }
}

\author{Namwoo Kang\thanks{Address all correspondence for other issues to this author.}
    \affiliation{
	Associate Professor\\
	Cho Chun Shik Graduate School of Mobility\\
	Korea Advanced Institute of Science and Technology\\
        Daejeon 34051, South Korea\\
    Email: nwkang@kaist.ac.kr
    }	
}

\begin{document}

\maketitle    

\begin{abstract}
\section*{Abstract}
{

\textit{Recent advances in artificial intelligence (AI) have impacted various fields, including mechanical engineering. However, the development of diverse, high-quality datasets for structural analysis remains a challenge. Traditional datasets, like the jet engine bracket dataset, are limited by small sample sizes, hindering the creation of robust surrogate models. This study introduces the DeepJEB dataset, generated through deep generative models and automated simulation pipelines, to address these limitations. DeepJEB offers comprehensive 3D geometries and corresponding structural analysis data.
Key experiments validated its effectiveness, showing significant improvements in surrogate model performance. Models trained on DeepJEB achieved up to a 23\% increase in the coefficient of determination and over a 70\% reduction in mean absolute percentage error (MAPE) compared to those trained on traditional datasets. These results underscore the superior generalization capabilities of DeepJEB.
By supporting advanced modeling techniques, such as graph neural networks (GNNs) and convolutional neural networks (CNNs), DeepJEB enables more accurate predictions in structural performance.
The DeepJEB dataset is publicly accessible at: \href{https://www.narnia.ai/dataset}{https://www.narnia.ai/dataset}.}
}
\end{abstract}



\section{Introduction}

The rapid advancements in artificial intelligence (AI) have revolutionized various domains, and the field of mechanical engineering is no exception. The significance of comprehensive and diverse datasets has become increasingly apparent as AI advances at an unprecedented rate. Generative models, particularly those based on 2D images and text, have already demonstrated their potential. Studies such as \cite{ramesh2022hierarchical, team2023gemini, achiam2023gpt, shen2024hugginggpt} have advanced into services that directly assist human life by training large models on high-quality datasets. These models have advanced from academic research to real-world applications, providing practical benefits to society. Inspired by these developments, researchers in the field of mechanical engineering are increasingly leveraging data-driven approaches. These approaches are being utilized to address complex problems previously considered intractable or bottlenecked.

AI has found extensive applications in two key areas within mechanical engineering: generative design and predictive analysis \cite{bappy2024exploring, patel2021artificial, tapeh2023artificial}. Generative AI has garnered significant attention as a potential solution to the scarcity of data in mechanical design. Moreover, generative AI can effectively address the limitations caused by insufficient real-world data by providing novel data augmentation techniques \cite{oh2019deep, regenwetter2022deep, shu20203d, wang2022ih}. Predictive analysis has emerged as a subcategory of data-driven surrogate modeling and reduced-order modeling research. These approaches aim to develop efficient and accurate models that can predict the behavior of complex engineering systems.

One of the significant advantages of AI models, particularly deep learning architectures, is their ability to learn from high-dimensional data. This capability sets these AI models apart from traditional data-driven methodologies, which often struggle to handle the complexity and dimensionality of real-world engineering problems \cite{belani2019requirements, goh2021review, zhang2021application, chang2022towards, cunningham2019investigation, du2022deep}. However, research in this area has been primarily limited to benchmark problems due to the need for additional available data. Studies on 3D data and field predictions, crucial for industrial applications, are particularly scarce. For this reason, when developing predictive models, it is common to define the problem using scalar values like minimum or maximum values as the objective \cite{li2019surrogate, li2022machine}. Although scalar values help in evaluating and optimizing shapes, they do not provide insights into the specific design regions that influence these evaluations. It is essential to identify which design domain led to this evaluation. In particular, since most scalar values are derived from field results, it is even more important to predict the entire field \cite{umetani2018learning, maurizi2022predicting, yang2022amgnet, rosset2023interactive}.

Addressing these challenges requires datasets that encompass field information, enabling comprehensive and insightful analyses in mechanical engineering applications \cite{regenwetter2022deep, kim2021knowledge}. Researchers and practitioners can develop data-driven models that optimize designs and provide an extensive understanding of the underlying phenomena governing system behavior by leveraging such datasets. This paradigm shift toward data-driven approaches has the potential to revolutionize the field of mechanical engineering, enabling the development of highly efficient, reliable, and innovative solutions to complex problems.

However, the availability of public datasets in the mechanical engineering field is limited compared with the actively researched computer science domain. In particular, 3D simulation data are scarce due to the high computational and time costs involved. Moreover, additional diverse 3D data benchmarks are necessary to support various types of analyses. Consequently, 3D simulation datasets with multiple analyses performed are rare, and the existing datasets require additional samples for effective machine learning.

Simulated jet engine bracket dataset (SimJEB) \cite{whalen2021simjeb} is one of the few publicly available 3D datasets in the engineering domain that provides boundary representation (B-Rep) files and structural analysis simulation data, including scalar and field data. However, the SimJEB dataset consists of only 381 samples, which must be increased to effectively train AI models. Furthermore, the structural analysis simulation label data contain outliers that significantly deviate from the data distribution. Additional data-cleaning processes are necessary to use this dataset for deep learning purposes, which may further reduce the number of usable samples.

To address these limitations, this work aims to utilize deep generative models to enhance the diversity of shapes and performance features in the SimJEB dataset, which serves as the foundation for our study. Our objective is to improve the quality and quantity of the data, resulting in a high-quality 3D dataset for engineering design applications. The proposed dataset, named DeepJEB, provides engineering data labeled through structural simulations and is intended to serve as a benchmark dataset for data-driven surrogate models in the field of structural analysis.

The main contributions of this work are as follows:
\begin{enumerate}

    \item We present a large-scale dataset containing high-resolution 3D designs of jet engine brackets, accompanied by their corresponding structural simulation results. This dataset comprises a total of 2138 designs, making it approximately 5.6 times larger than SimJEB, the previously largest publicly available dataset for jet engine brackets. Moreover, this dataset provides information for estimating engineering performance based on various structural designs.
    
    \item Our dataset utilizes second-order tetrahedral elements in 3D structural simulations compared with SimJEB, providing more accurate simulation values and detailed field data based on an average of 209,000 nodal values.
    
    \item In addition to the four linear static load cases (horizontal, vertical, diagonal, and torsional) performed in the baseline dataset, we introduce supplementary engineering performance metrics not available in SimJEB, such as natural frequencies, mode shapes obtained through normal mode analysis, and inertia tensors, which enable the evaluation of dynamic performance.

    \item We provide multi-view images applicable to multi-view models, a popular research topic in the 3D graphics domain. Our dataset, unlike SimJEB, includes these multi-view images, facilitating advanced modeling techniques and analysis that are increasingly relevant in contemporary 3D graphics research.
    
    \item We demonstrate the potential of deep generative models to expand the design space of jet engine brackets and propose a simulation pipeline methodology for labeling engineering performance based on the synthesized data.
    
    \item We create a training dataset suitable for deep learning by expanding the dataset size and demonstrate the influence of synthetic data on improving surrogate model performance through a case study. The case study shows a 22.8\% improvement in R² score compared to the baseline dataset, indicating enhanced predictive accuracy.
    
    \item We conduct a comprehensive analysis of the design and performance space to pre-label an appropriate test dataset that can serve as a benchmark for deep learning training.
    
\end{enumerate}

\begin{table}[ht]
\centering

\begin{tabular}{|l|p{5cm}|}
\hline
\textbf{Data Type} & \textbf{Description} \\ \hline
Surface Mesh       & Tessellated surface mesh data in STL format.\\ \hline
Volume Mesh        & Second-order tetrahedral mesh data in VTK format.\\ \hline
B-Rep              & Boundary representation data in STEP format.\\ \hline
Solver Deck        & Input files for numerical analysis solvers in FEM format.\\ \hline
Scalar Data        & Geometric and analysis information in CSV format, including center of gravity, mass, volume, maximum displacement, maximum stress, and natural frequencies. \\ \hline
Field Data         & Detailed simulation results at mesh nodes in CSV format, encompassing the $x$, $y$ and $z$ coordinates, and resultant displacement, signed von Mises stress, and normal mode shapes.\\ \hline
Field Mesh         & Hierarchical data format
file combining volume mesh and nodal data in H5 format.\\ \hline
Image& Images of the bracket from various angles in PNG format.\\ \hline
 Metadata&Train-test split labels in JSON format\\\hline
\end{tabular}
\caption{DeepJEB dataset overview}
\label{tab:deepjeb_overview}

\end{table}

The DeepJEB dataset is categorized into geometric data and analysis data. Geometric data are available as surface mesh, B-Rep, and volume mesh data (Table~\ref{tab:deepjeb_overview}). Analysis data encompasses scalar and field data for each sample and analysis solver input files. We also provide H5 files that combine the volume mesh and field data to facilitate efficient processing and mapping of field data with geometric data.

Scalar data include geometric information, such as the center of gravity, mass, and volume of each data sample and analysis information, such as maximum displacement, maximum stress, and natural frequencies. Field data encompass detailed simulation results at each mesh node, including the $x$, $y$ and $z$ coordinates, corresponding displacement, and stress values.

We developed a dataset comprising 2138 3D geometric samples and their corresponding structural analysis data utilizing deep learning techniques. We also provide signed von Mises stress data, which allow for determining tension/compression based on the sign. These specific forms of data ensure that surrogate models can be evaluated for their performance and robustness in a wide range of scenarios.

Furthermore, we provide a dataset of 26 multi-view images for each bracket, including 8 distinct azimuth angles, 3 elevation angles, as well as top and bottom views. We include train–test split labels in JSON format to facilitate the use of the dataset as a benchmark, ensuring a uniform distribution across the design and performance spaces.

The structure of our paper is organized as follows: 
Section~\ref{Literature Review} reviews related work, including benchmark datasets and the use of deep generative models in the field. 
Section~\ref{DeepJEB Dataset Creation} describes the methodologies used for data collection, generation, geometric filtering, and the automation of analysis processes. 
Section~\ref{Dataset Validation} details the validation processes utilized to ensure data quality and benchmark suitability, comparing the design spaces between SimJEB and DeepJEB datasets. 
Section~\ref{Case Study} presents a case study demonstrating the application of the synthetic dataset in surrogate modeling, comparing the performance of models built with the SimJEB and DeepJEB datasets. 
Section~\ref{Conclusion} summarizes the findings and contributions of this work and discusses implications for future research. Finally, Section~\ref{Licensing} provides information on licensing, attributions, and access to the dataset and research tools used.

\section{Literature Review}
\label{Literature Review}

\subsection{Benchmark Dataset}

The rapid advancements in computer science have led to a proliferation of studies focused on 3D models, resulting in an increased availability of datasets for related research. In contrast to 2D images, which are defined as structured grids, 3D data can be represented in various forms, such as voxels (discretized volumes analogous to images), multi-view images (captured from different angles), point clouds (surfaces discretized into points), and more.

Computer-aided design (CAD) data, consisting of the commands used to create 3D models, also falls under the category of 3D data. Research endeavors, such as Fusion360 Gallery \cite{willis2021fusion} and DeepCAD \cite{wu2021deepcad}, have aimed to generate data based on these command sequences. Mesh data can be readily converted into point clouds, voxels, and multi-view representations.

Several benchmark datasets, such as ShapeNet \cite{chang2015shapenet}, ABC \cite{koch2019abc}, and ModelNet \cite{wu20153d}, have been established to represent 3D shapes using mesh-based approaches. However, these datasets were primarily constructed by research communities with a background in computer graphics, resulting in a higher proportion of everyday objects and household items compared to engineering components that could be utilized in a mechanical engineering context.

Among these datasets, ShapeNet has gained popularity due to its diverse range of categories and a large number of data samples within each category, resulting in studies leveraging this dataset. Previous works \cite{song2023surrogate} and \cite{cunningham2019investigation} utilized the vehicle and airplane categories from ShapeNet to perform computational fluid dynamics (CFD) analyses and develop data-driven surrogate models and optimization techniques.

The MCB dataset \cite{kim2020large}, which focuses on mechanical parts such as gears, brackets, and linkages, offers high potential for practical applications. However, the dataset contains a significant number of component-level data and was collected from various sources, such as Traceparts, GrabCAD, and 3D Warehouse, resulting in inconsistencies in alignment and scaling, which limits its usability.

A notable drawback of existing benchmark datasets, including ShapeNet, is the lack of labels related to engineering performance, which limits their applicability to relevant engineering problems. Performance data is necessary for the development of data-driven models that can predict and optimize the behavior of 3D shapes in an engineering context.

Efforts have been made in the engineering field to develop large-scale 3D datasets specifically designed for this area to address these limitations. These efforts aim to address the shortcomings of existing datasets and provide a foundation for data-driven approaches in engineering applications. Notable examples include the FRAMED dataset \cite{regenwetter4132282framed}, which consists of 4500 parametric bike frame shape data coupled with finite element method (FEM) results, and the Ship-D dataset \cite{bagazinski2023ship}, which contains ship-hull data defined by 45 parameters along with 32 wave drag coefficients. In a similar vein, a large-scale multimodal car dataset, DrivAerNet++ \cite{elrefaie2024drivaernetlargescalemultimodalcar}, was introduced to integrate CFD simulations with deep learning benchmarks to enhance vehicle aerodynamics design and optimization. This dataset exemplifies the potential of combining high-fidelity simulations with machine learning applications to improve predictive modeling and performance analysis in engineering.

SimJEB \cite{whalen2021simjeb} is a public dataset that includes shape data, scalar values, and FEM data. This dataset utilizes 381 hand-designed CAD models submitted to the GE jet engine bracket challenge. SimJEB provides scalar labels, such as maximum displacement in the $x$, $y$, and $z$ directions, and maximum von Mises stress for four load cases. Additionally, SimJEB offers field data for the nodal points from the finite element analysis (FEA). However, SimJEB has a limitation in terms of the absolute quantity of data, making it challenging to obtain a sufficient dataset for training and validating deep learning models.

\subsection{Deep Generative Models for Engineering Applications}

The advancement of deep learning techniques and the growing interest in generative models have resulted in numerous efforts to utilize these models to generate engineering data. In particular, numerous studies have focused on 2D image-based approaches.

In mechanical design, research on topology optimization has been actively conducted even before the emergence of generative models. Studies \cite{giannone2024aligning, maze2023diffusion, nie2021topologygan, oh2019deep}, and various reviews \cite{shin2023topology, maksum2022computational, mukherjee2021accelerating} have explored the integration of topology optimization with generative models.

TopOpNet \cite{jang2022generative} proposed a methodology that leverages reinforcement learning to generate many diverse data samples within a short inference time. Reference \cite{oh2019deep} introduced a complete process for training generative models using 2D image data generated through topology optimization methods. This research was further extended by \cite{yoo2021integrating}, who developed a framework for generating and analyzing 3D data.

In the domain of CFD, researchers have been conducting studies on data-driven generative models using existing benchmark datasets, such as airfoils, and CFD simulators, such as X-foil \cite{drela1989xfoil}. PaDGAN \cite{chen2021padgan} presented a generative model that utilizes Bezier curve techniques, taking advantage of the smooth surface characteristics and parameterized data of airfoils. Moreover, PaDGAN utilized a determinantal point processes (DPP) loss function to promote the generation of high-quality samples while covering a diverse data space.

PcDGAN \cite{heyrani2021pcdgan} is an improved generative model that aims to generate new designs that satisfy specific performance requirements with an enhanced ability to model continuous conditions. This model incorporates concepts, such as performance-conditioned DPP loss, singular vicinal loss, and conditional batch normalization, for continuous labels to ensure continuity and diversity from a performance perspective.

Furthermore, MO-PaDGAN \cite{chen2021mo} has been proposed as an extension of the PaDGAN, incorporating DPP-based loss functions suitable for multi-objective problems, enabling the discovery of superior Pareto fronts.

Synthetic data have gained significant attention in engineering design as a crucial solution to address the limitations of real-world datasets in terms of scale and availability. For example, Saha et al. \cite{Saha2021Exploiting} explored the use of autoencoders and variational autoencoders to predict the aerodynamic performance of 3D car designs. By leveraging latent representations learned from 3D point cloud data, their approach effectively integrated deep generative models with CFD simulations to enable rapid performance predictions and optimize design processes. This integration of generative models and simulation highlights their potential to enhance data-driven approaches in engineering. Similarly, CarHoods10k \cite{wollstadt2022carhoods10k} provided a synthetic dataset that includes over 10,000 data samples and corresponding performance data. This synthetic dataset was generated by identifying and parameterizing features from CAD models of automotive hoods, which were combined with basic shapes and feature patterns to produce synthetic data. The synthesized data leveraged geometric deep learning techniques to predict performance values within a low-dimensional latent space instead of relying on traditional parametric methods. Additionally, these techniques facilitated the generation of new samples, thereby enabling design optimization and the creation of novel geometric configurations. Furthermore, research efforts, such as the DATED \cite{picard2023dated} study, have focused on establishing guidelines for generating high-quality synthetic datasets in the engineering design domain.

\subsection{3D Generative Model}

Although 2D data offer advantages in training and have numerous benchmark datasets that facilitate research, their validity has limitations, as real-world mechanical components and their analyses are primarily performed in three dimensions. 3D data pose challenges due to the limited availability of usable datasets and the complexities involved in representing and processing 3D shapes \cite{regenwetter2022deep}.

The representation of 3D data varies depending on the problem definition \cite{guo2020deep, chaudhuri2020learning, park2019deepsdf, chabra2020deep, mildenhall2021nerf}. The main categories of 3D data representation include voxels, point clouds, meshes, and implicit functions. For instance, Umetani \cite{Umetani2017Exploring} explored the use of autoencoder networks for converting unstructured triangle meshes into consistent topologies, providing an efficient method for parameterizing 3D shapes in machine learning applications. This technique allows for the exploration and synthesis of new shapes through a compact latent space, demonstrating the potential of generative models in creating diverse 3D datasets. Voxels discretize 3D volumes into a grid, enabling the use of convolutional neural networks and similar techniques \cite{maturana2015voxnet, li2021survey}. However, voxel representations can lead to memory inefficiencies and higher computational costs due to unnecessary regions, and the resolution can often result in less detailed and blunt shapes. Point clouds represent 3D surfaces using a set of 3D points, capturing detailed shapes and providing a global representation \cite{qi2017pointnet, qi2017pointnet++}. Despite this, point clouds lack the ability to capture connectivity or topological information between points, making watertight mesh generation challenging. Meshes discretize surfaces by connecting vertices to form faces, offering a compact representation and compatibility with graph-based approaches \cite{nash2020polygen, gao2022get3d, huang2021arapreg, muralikrishnan2022glass}. However, deforming meshes can be problematic as changing topology is difficult, and self-intersections can occur, rendering them less suitable for mechanical designs requiring topological flexibility.

Due to these limitations, implicit functions are increasingly used as a representation for generative models, especially for mechanical components \cite{peng1999pde, osher2004level}. The signed distance function (SDF) is commonly utilized for solid shapes with a clear distinction between the interior and exterior. An SDF represents the shortest distance from an arbitrary point to the shape's surface and is determined by whether the point is inside or outside the shape, indicated by its sign.

Recent advancements in this area include the development of models like Shap·E \cite{jun2023shape}, which generate conditional 3D implicit functions using a novel two-stage training process involving an encoder and a conditional diffusion model. This approach enables the direct generation of implicit neural representations, which can be rendered as both textured meshes and neural radiance fields, offering significant improvements in efficiency and flexibility over traditional explicit models. These advancements underscore the growing potential of implicit neural representations in creating diverse and high-quality 3D geometry efficiently.

Implicit neural representations, particularly SDF, have gained significant attention in 3D generative models due to their ability to represent shapes with high fidelity, handle topological changes, and generate novel and feasible shapes. These representations have shown promising results in various domains and have the potential to revolutionize the way 3D data is represented and processed in engineering applications.

Since DeepSDF \cite{park2019deepsdf}, the pioneering approach to representing shapes as signed distance functions and implicitly learning them, numerous models have been developed for shape manipulation. Among these, DualSDF \cite{hao2020dualsdf} introduces an additional coarse network to facilitate shape manipulation, while A-SDF \cite{mu2021sdf} disentangles the latent space specifically for articulated objects. These models share a common feature: an auto-decoder structure without an encoder. In addition, these models take coordinates and latent code as inputs, optimizing the latent code within the model via backpropagation.

\section{DeepJEB Dataset Creation}
\label{DeepJEB Dataset Creation}

Developing the DeepJEB dataset represents a comprehensive effort to create a high-quality, expansive dataset tailored for engineering design and analysis applications. This section details the processes and methodologies used to transform the SimJEB dataset into the enhanced DeepJEB dataset.

\subsection*{Introduction of Baseline Dataset}
The DeepJEB dataset, derived from the SimJEB 3D shape dataset, originates from the GE jet engine bracket CAD challenge. This dataset, comprising various CAD models specifically designed based on structural analysis of jet engine brackets, is a valuable resource for engineering applications. In creating DeepJEB, the geometry data from SimJEB were utilized to create diverse bracket designs. FEM simulations were then performed on the derived synthetic data to enrich the dataset with necessary engineering analysis data.

\begin{figure}[h!]
    \centering
    \begin{subfigure}[b]{3.34in}
        \includegraphics[width=\textwidth]{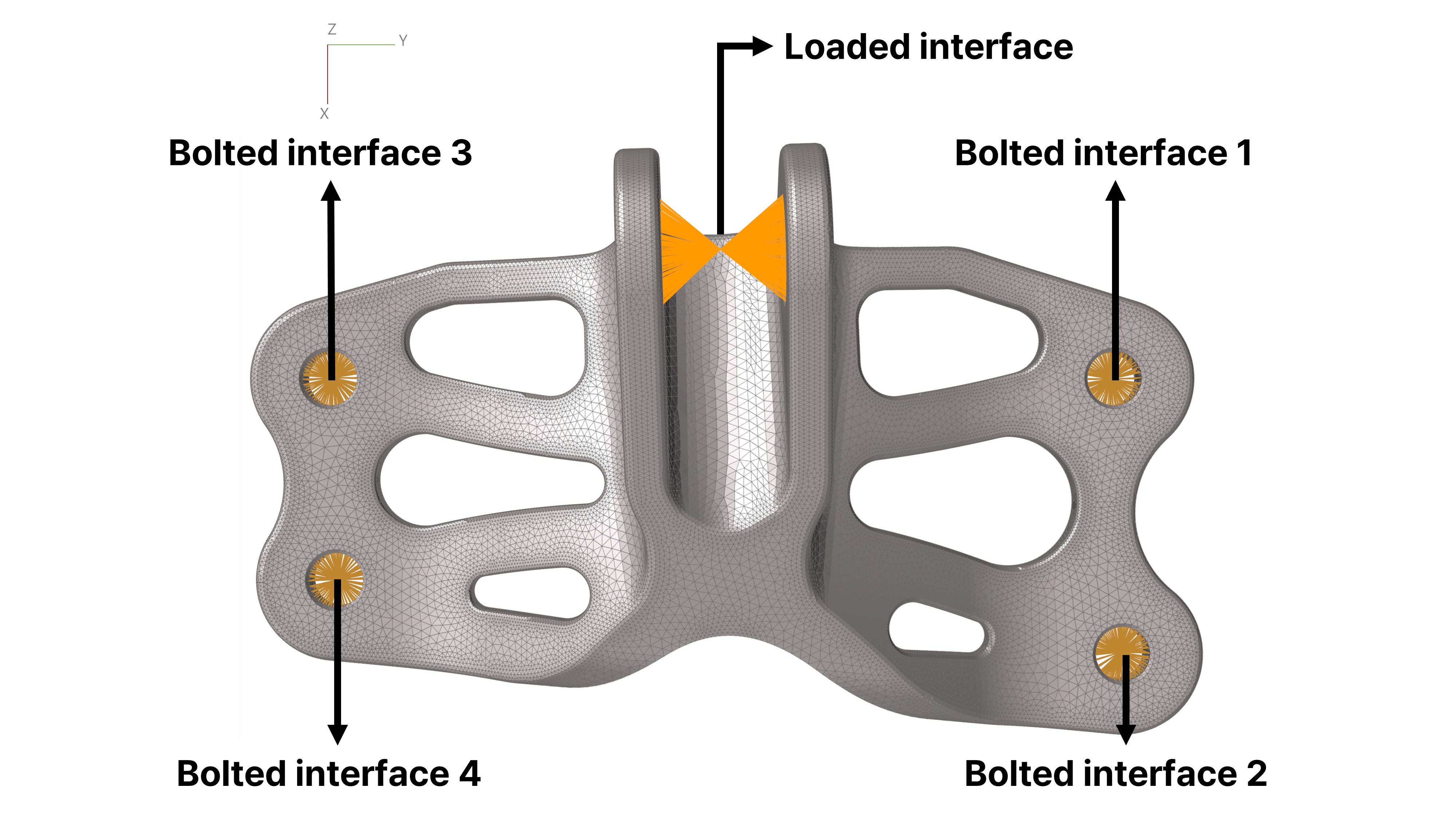}
        \caption{SimJEB RBE boundary interface}
        \label{fig:rbe_bc}
    \end{subfigure}
    \hfill 
    
    \begin{subfigure}[b]{0.40\textwidth} 
        \includegraphics[width=\textwidth]{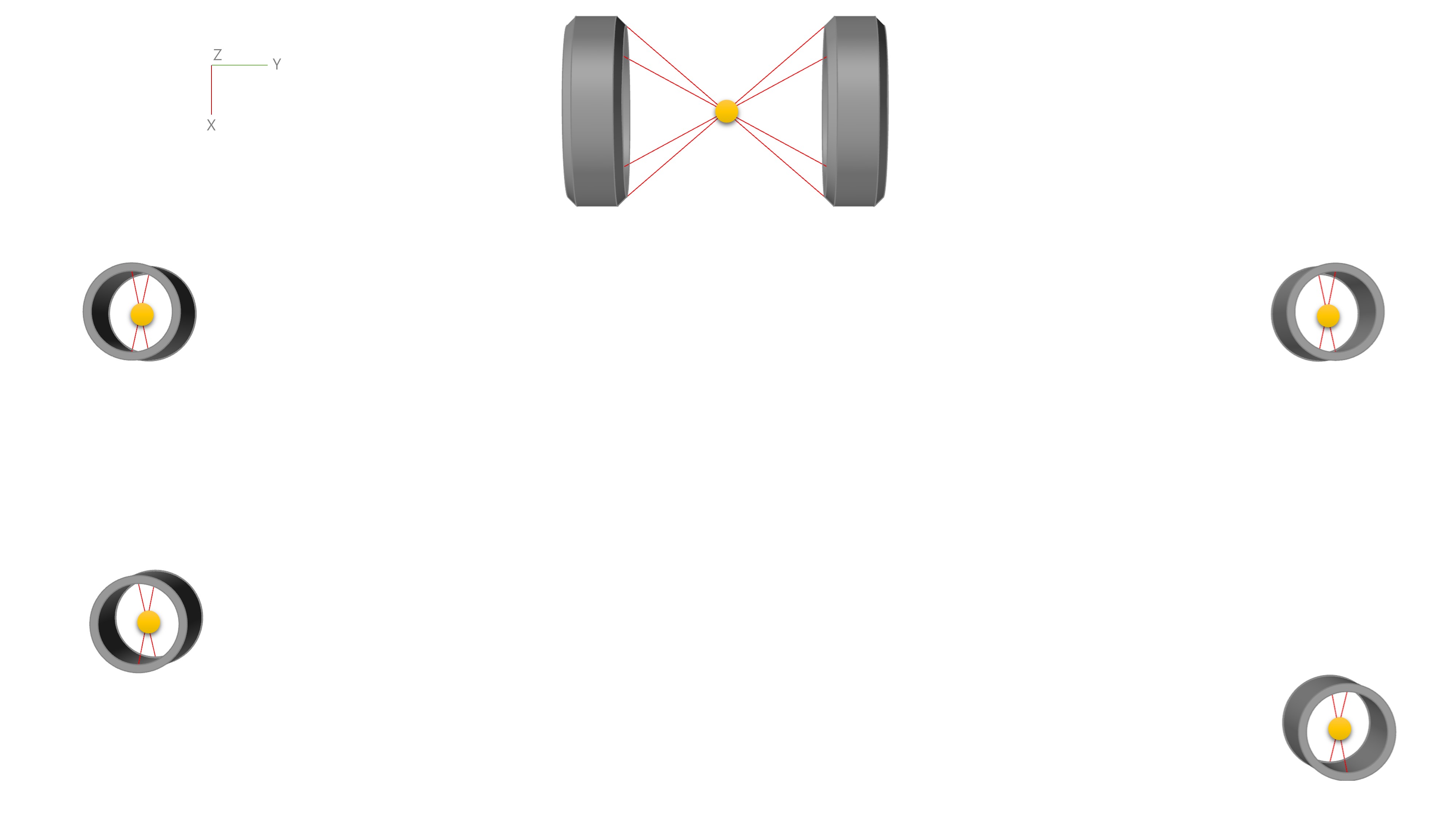}
        \caption{Predefined center points}
        \label{fig:rbe_center}
    \end{subfigure}
    \caption{RBE boundary conditions and predefined center points}
    \label{fig:rbe}
\end{figure}

The boundary conditions in the baseline dataset are derived from the GE Jet Engine Bracket CAD challenge interfaces. These conditions are integrated into rigid body elements (RBE) to enhance computational efficiency. Figure~\ref{fig:rbe_bc} illustrates the categorization of the boundary conditions into loaded and bolted interfaces. The loaded interfaces are connected using multi-point constraints (MPC) and are modeled as interpolated RBE, specifically RBE3. An RBE3 is used to distribute loads or constraints across multiple nodes; it connects several dependent nodes to an independent node and averages the dependent nodes' movements to match that of the independent node. This approach is effective in simulating flexible connections and ensures that the simulation efficiently represents the load distribution and structural response. For example, points near the loaded interface are connected via an MPC and modeled as an RBE3, with the center point of this element subjected to predefined loads. Meanwhile, the bolted interfaces are also connected using MPC but are modeled as RBE2. An RBE2 connects a set of independent nodes to a single dependent node, ensuring that the dependent node’s movement represents the average movement of the independent nodes, which is typically used for modeling rigid connections. For instance, bolted interfaces 1 to 4 use an RBE2 where the points near these interfaces are fixed, providing a realistic simulation of the bracket’s attachment points to the engine structure. 

\subsection{Deep Generative Model-based Geometry Generation}

\subsubsection{Seed Data Selection}
Seed data is meticulously selected from the baseline dataset to ensure the stability and reliability of the synthetic data. The selection process aims to eliminate samples that may introduce inconsistencies and maintain uniformity in boundary conditions during the automated simulation stage. Ensuring consistency in boundary conditions is crucial for maintaining geometric stability and preventing errors in the simulation pipeline, which could otherwise arise from variations at boundary sections.

To this end, samples with geometric issues, such as those prone to noise, cavities, or fragile structures, were removed. In particular, samples with cavities were excluded because they could pose significant challenges during the subsequent interpolation process. Cavity samples, especially those with thin-walled structures, often exhibit geometric patterns that differ significantly from those without such features. This disparity can lead to inconsistencies when generating new shapes, as the interpolation process relies on smooth transitions between data points. The presence of cavities introduces the risk of creating non-uniform distributions and geometric noise, which could compromise the quality of the synthetic data and reduce its reliability. Moreover, geometric noise that arises within interpolated shapes, particularly from cavities, can be difficult to detect in advance. Removing these samples ensures the overall geometric stability and consistency of the dataset.

Furthermore, alignment errors between the predefined boundary center points (illustrated in Fig.~\ref{fig:rbe_center}) and the sample interfaces are calculated. The distance between the RBE center point of each interface and the RBE center point of the predefined boundary design is calculated according to Eq.~(\ref{eq:bal}) to quantify the boundary alignment error.

\begin{align}
    &{\text{Sample RBE center}_{\text{interface}}}(x_i, y_i, z_i) \notag \\
    &\text{Predefined center}_{\text{interface}}(x_i, y_i, z_i) \notag \\
    \tiny&\text{Boundary alignment error} \notag \\
    &\space= \scriptsize \sum_{\text{i}=1}^{5} \Vert\text{sample RBE center}_{\text{i}} - \text{predefined center}_{\text{i}}\Vert \tag{1} \label{eq:bal}
\end{align}
\normalsize

The cumulative error across all five interfaces determines the overall alignment error. The average alignment error was calculated as 6.8, and a conservative threshold of 3.0 was set for the final alignment error.

\begin{figure}[h!]
    \centering
    \begin{subfigure}[t]{0.45\textwidth} 
        \includegraphics[width=\textwidth]{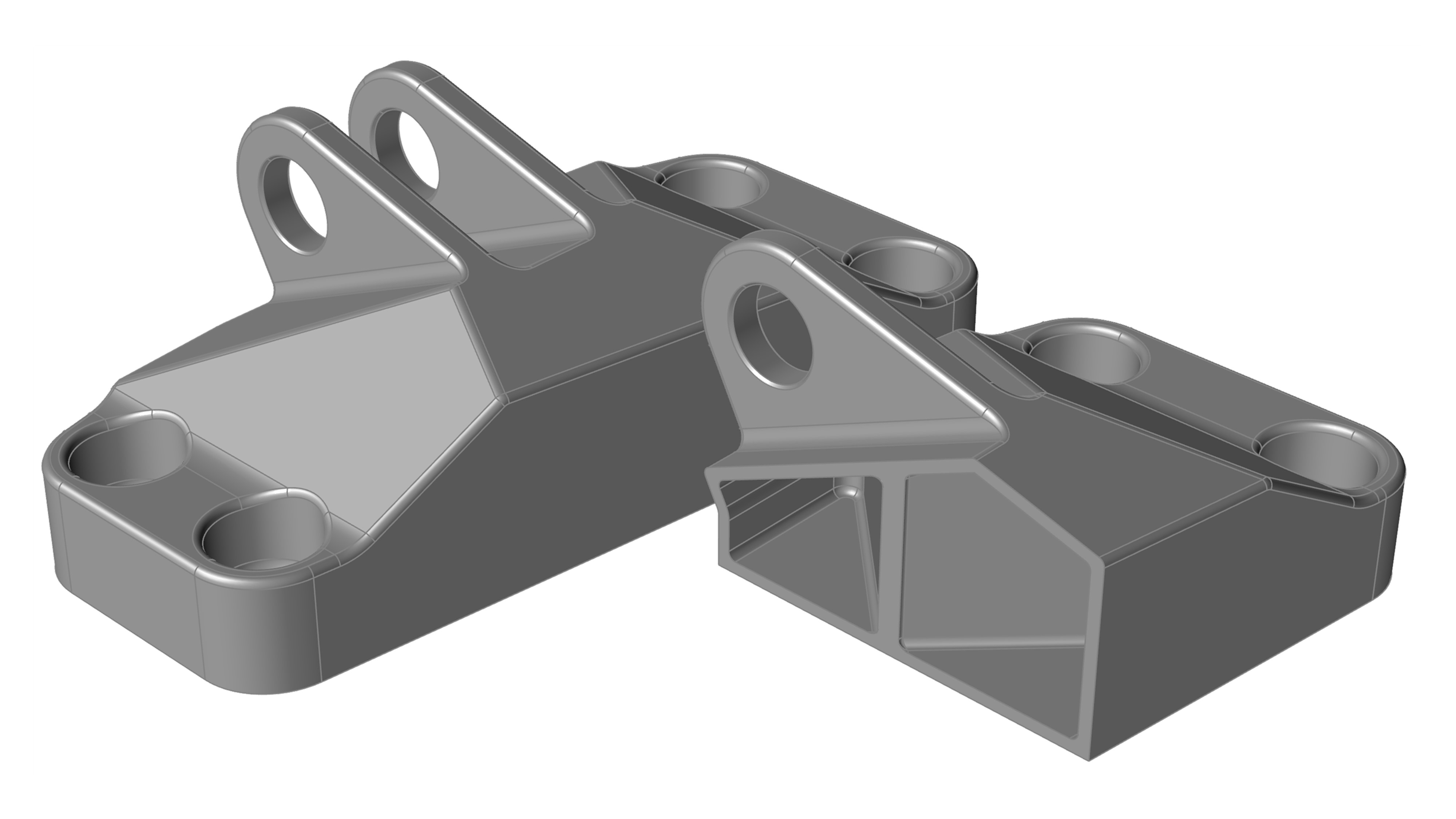}
        \caption{Samples with cavities}
        \vspace{1cm}
        \label{fig:cavity}
    \end{subfigure}
    \begin{subfigure}[tbh]{3.34in}
        \includegraphics[width=\textwidth]{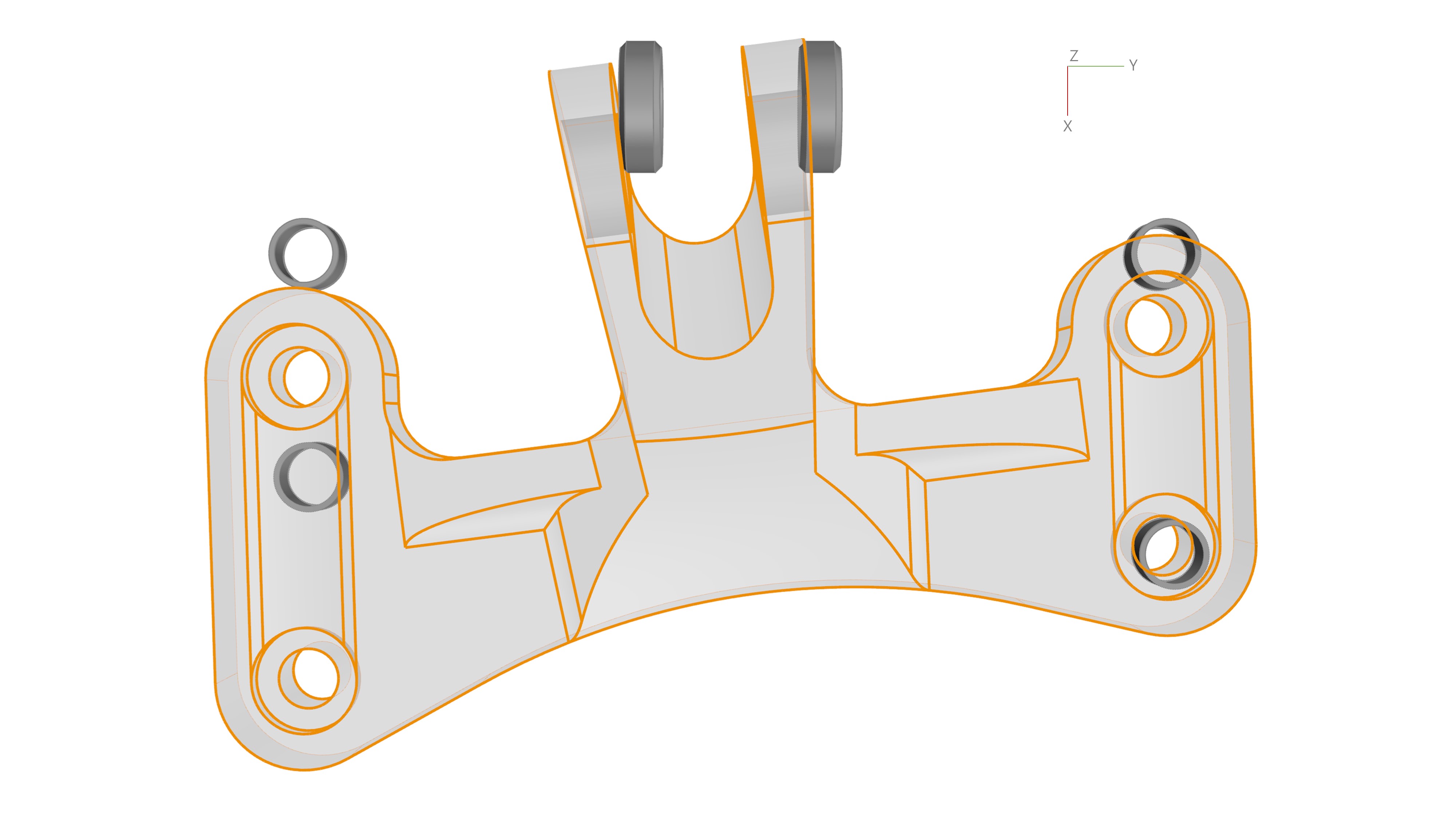}
        \caption{Samples with boundary alignment errors}
        \label{fig:bc_error}
    \end{subfigure}
    \caption{Examples of unselected samples due to geometric issues}
    \label{fig:unselected_geometry}
\end{figure}

For the generated geometries, the RBE deviation metric was not calculated separately because the boundary conditions were defined based on predefined geometries. These predefined geometries, which include the boundary conditions, were combined with the generated geometries using boolean operations. As a result, the RBE centers and the boundary points are precisely aligned, leading to a theoretical deviation of zero. This ensures that there is no discrepancy or deviation in the alignment, eliminating the need for any additional alignment adjustments for the generated data.

The selection process resulted in the removal of 118 samples. Examples of these removed samples are illustrated in Fig.~\ref{fig:unselected_geometry}. Consequently, 263 samples with well-aligned boundary conditions and stable geometries were selected as seed data. This selection process was carried out to ensure that the seed data used for generating new geometries were of high quality, thereby enhancing the reliability of the synthetic data generated in subsequent steps.

\subsubsection{Data Generation Process}

In this study, we adopted the implicit neural representation methodology. This approach is particularly practical in modeling complex data structures and projecting high-dimensional data into lower-dimensional spaces, which is essential for handling the intricate geometries and diverse topologies present in the baseline dataset. The primary objective of our data generation process is to augment the baseline dataset, which contains 263 bracket seed data samples after the selection process. Despite the reduction in the number of usable samples, meaningful data that maintain quality must be generated while enhancing the diversity and reliability of the dataset.

\begin{figure} 
\centerline{\includegraphics[width=3.34in]{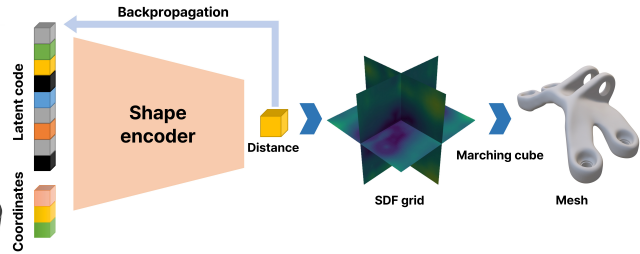}}
\caption{Auto-decoder architecture}
\label{fig:gen_arch}
\end{figure}

One of the significant challenges in this endeavor is constructing a meaningful latent space with the limited amount of data available. Methodologies that can learn complex patterns and perform compelling inferences from sparse data must be developed. This task requires approaches to maximize each data point’s utility and generate synthetic data that accurately represent the underlying distributions and variations in the original dataset. Accordingly, this study utilizes the implicit neural representation methodology, which introduces a new approach to handling high-dimensional data efficiently. This methodology provides a foundation for achieving high performance even with a limited dataset. The overall structure of the model is illustrated in Fig.~\ref{fig:gen_arch}. The architecture consists of eight linear layers, each with 512 dimensions, followed by a 128-dimensional latent code. Weight normalization and ReLU activation functions are applied between each linear layer, with the final output passing through a Tanh activation function. The model is trained using the Adam optimizer with a learning rate of 5e-4. The loss function ($L_{AD}$) used for training the generative model consists of two components: an L1 loss and a regularization loss, as defined as Eq.~(\ref{eqn:loss_ad}). 
\begin{equation}
\label{eqn:loss_ad}
\tag{2}
L_{AD} = \sum_{i = 1}^{N}(L(f_{\theta}(z_i,x_i), s_i) + ||z||^2)
\end{equation}
The L1 loss measures the difference between the output of $f_{\theta}$, which takes predefined coordinates $x_i$ and the corresponding latent code $z_i$ as inputs, and the signed distance function $s_i$ at the location $x_i$. Additionally, the regularization loss ensures that the entire latent code is normalized by applying the L2 norm. The total loss is the sum of these two terms.

The methodology implementation steps include utilizing implicit neural representation to model the complex geometries of the jet engine brackets. This representation efficiently handles high-dimensional data and provides a scalable solution for data augmentation. The next step is constructing a latent space that captures the bracket design's essential features and variations. The latent space must be robust enough to facilitate the generation of diverse and meaningful synthetic data. Finally, the implicit neural representation methodology are applied to the selected seed data to generate new bracket designs. The generated data is then validated to ensure they meet our study's quality and diversity criteria.
After the data generation process, we obtained 4833 synthetic bracket designs.

\subsubsection{Synthetic Data Shape Filtering}

The DeepJEB dataset comprises synthetic data generated from various topologies and geometric variations. Given the complexity and variety of shapes in the dataset, some generated geometries may exhibit irregularities, such as tears or disconnected regions. Although these issues are a trade-off for achieving high diversity from a limited initial dataset, they necessitate a rigorous filtering process to ensure the dataset's practical usability. The filtering process is essential to eliminate synthetic geometries that are torn or disconnected, which are unsuitable for practical applications. The integrity and quality of these shapes must be ensured to maintain the dataset’s reliability and effectiveness in the subsequent analyses and applications.

Various mesh quality metrics were reviewed to evaluate the quality of generated meshes, including minimum skewness, minimum angle, and maximum angle \cite{gokhale2008practical}. These metrics are useful for identifying local geometric issues that can degrade the quality of synthetic data. However, through experimental evaluations, the minimum Jacobian determinant (min jac) metric emerged as the most effective for this purpose.

The min jac metric assesses the collapse degree of each mesh element, providing a robust indicator of geometric abnormalities. High min jac values correspond to stable and well-formed meshes, while low values signal potential issues such as element collapse or geometric inconsistencies \cite{knupp2001algebraic}. By focusing on extremal metrics rather than mean values, this approach allows for the detection of localized issues that could compromise the overall integrity of the mesh.

The synthetic data used in this study were initially defined in the Signed Distance Field (SDF) and then converted to a surface mesh using the marching cubes algorithm \cite{lorensen1998marching}. The quality of the resulting surface mesh was evaluated using the min jac metric. Despite the high geometric fidelity achievable through the marching cubes algorithm, some meshes exhibited poor quality in terms of angles and connectivity, highlighting the critical role of the min jac metric in assessing the quality of the surface mesh and identifying potential geometric inconsistencies.

The filtering process involves several steps to ensure the quality of the generated meshes. First, the synthetic data are converted from the SDF field to surface meshes by using the marching cubes algorithm. This algorithm reconstructs surface meshes that can be evaluated for geometric fidelity. Next, the quality of each generated mesh is assessed using the min jac metric. This metric evaluates the geometric integrity of the mesh, focusing on detecting abnormal shapes that may not be suitable for simulation or practical use. A threshold for the min jac metric is implemented to filter out substandard geometries. Meshes that do not meet the quality standards defined by the min jac metric are excluded from the dataset.

The min jac metric is crucial for the filtering process as it directly evaluates the geometric quality of the mesh. Unlike other metrics that may focus on mesh quality in terms of aspect ratios or element sizes, min jac specifically targets geometric stability and connectivity, which are vital for ensuring the dataset's practical usability. By filtering out meshes that fail to meet the min jac threshold, we ensure that only high-quality, well-formed geometries are included in the DeepJEB dataset. This process enhances the dataset's reliability and applicability.

The synthetic data shape filtering process filtered out 2096 data points, resulting in a final set of 2737 well-formed bracket designs.

\subsection{Automated Engineering Simulation pipeline}

The DeepJEB dataset creation process involves establishing a fully automated FEM simulation process using the filtered synthetic 3D bracket geometries, such as FEM input geometries. This process is built using Altair Inspire and SimLab \cite{altairAccelerateSimulationdriven, altairMultiphysicsWorkflows} to create an automated FEM simulation pipeline designed to effectively label engineering performance data.

The mesh specifications play a pivotal role in ensuring the accuracy and reliability of the simulation results. The mesh used in the simulations consists of second-order tetrahedral elements. Each second-order tetrahedral element includes additional mid-nodes, resulting in 10 calculation points per element. This mechanism enhances the accuracy of stress analysis by providing more detailed and precise stress distribution within each element, although it increases the computational time. The average element size is set to 2 mm, balancing computational efficiency with simulation accuracy. The bracket geometries in the dataset have maximum dimensions of approximately 186.42 mm in length, 111.61 mm in width, and 65.81 mm in height. Given these dimensions, the 2 mm mesh size is sufficient to capture the geometric details of the brackets while maintaining computational efficiency. The bracket material is set as Ti–6Al–4V, known for its specific properties: an elastic modulus ($E$) of 113.8 GPa, a Poisson's ratio ($\nu$) of 0.342, and a density ($\rho$) of $4.47\times10^{-3}$ g/mm³.

\begin{figure}[h!]
    \centering
    \begin{subfigure}[tbh]{3.34in}
        \includegraphics[width=\textwidth]{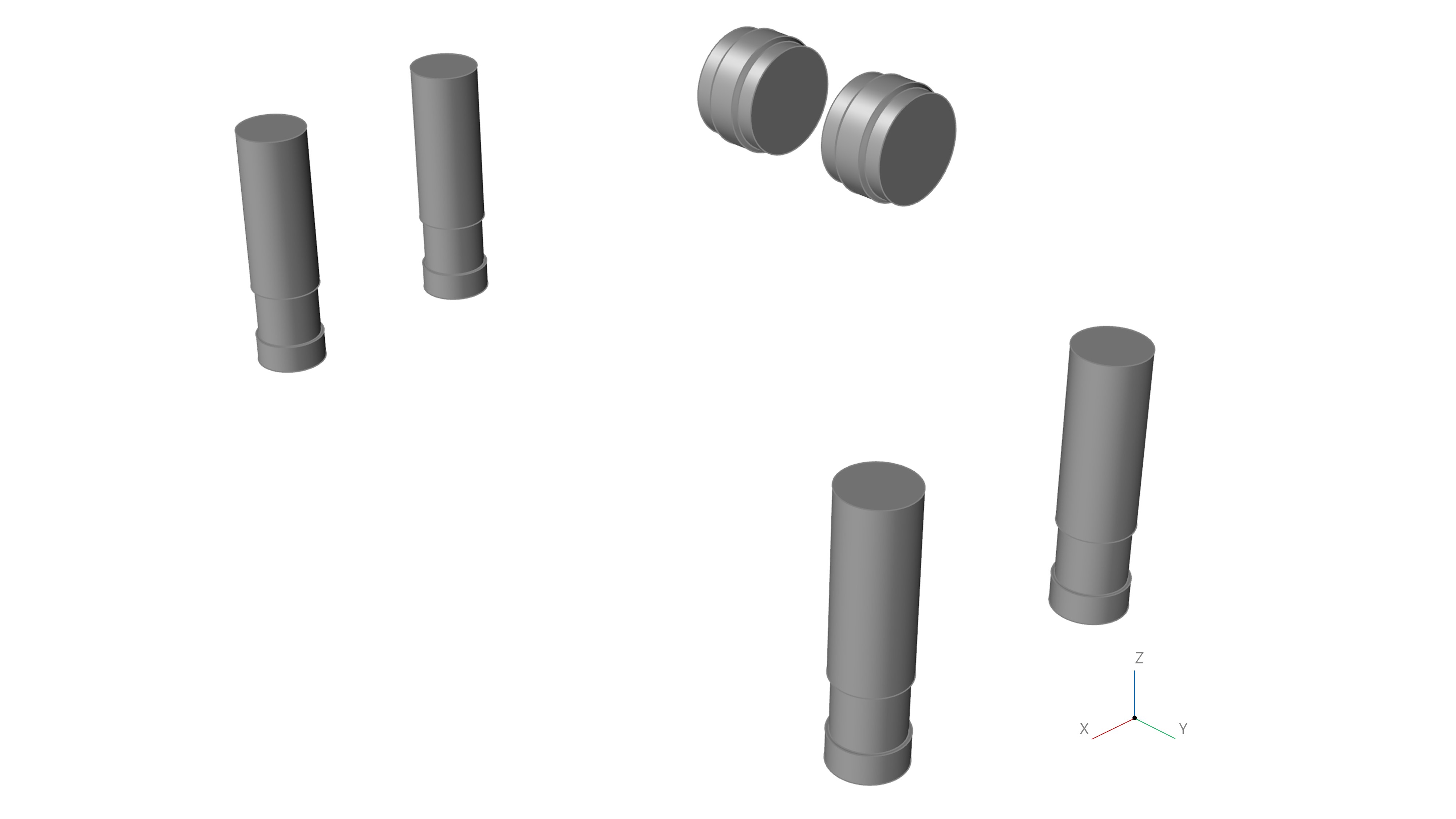}
        \caption{Boundary cleanup geometry}
        \vspace{1cm}
        \label{fig:subtract_geo}
    \end{subfigure}
    \begin{subfigure}[tbh]{3.34in}
        \includegraphics[width=\textwidth]{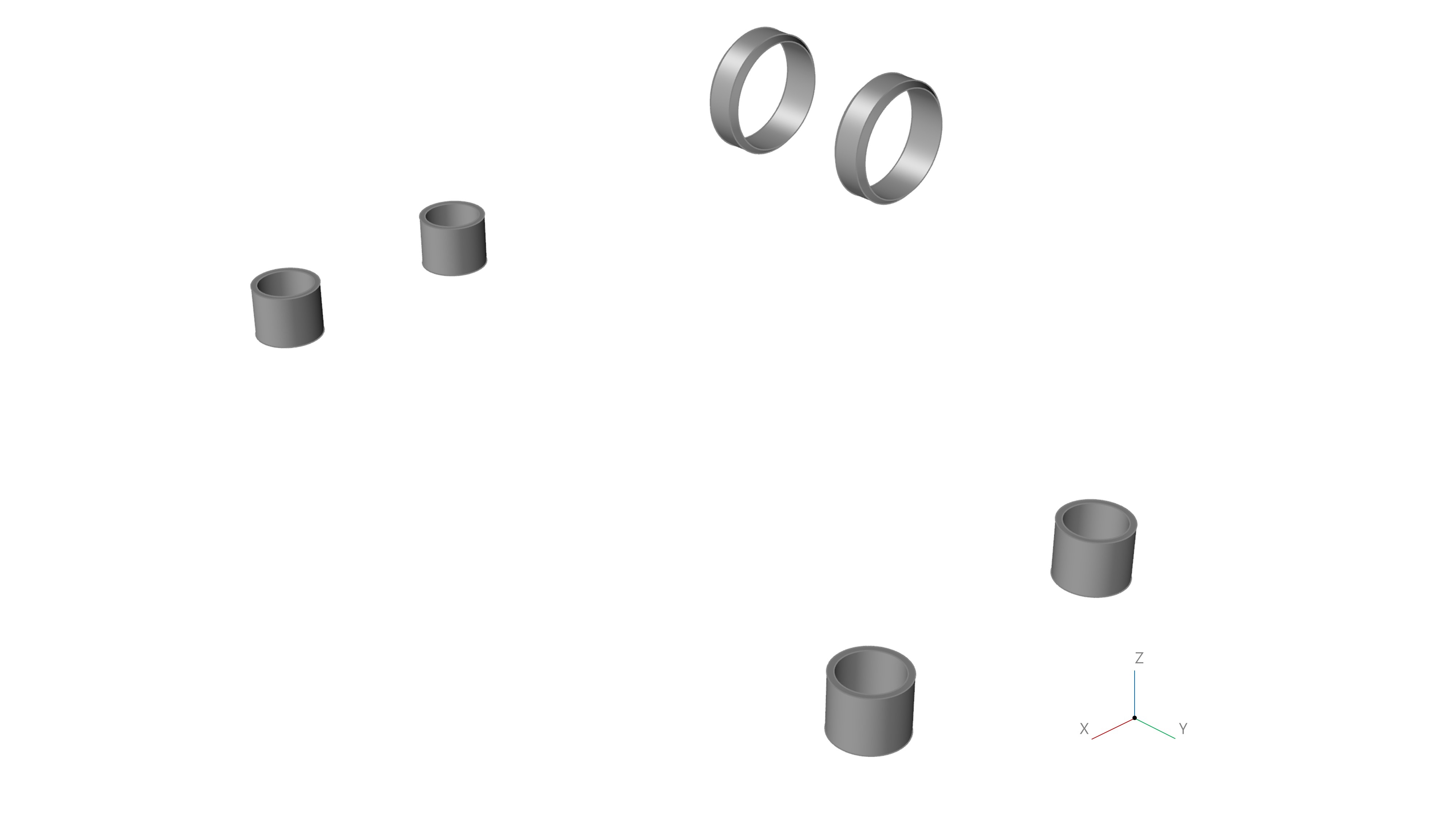}
        \caption{Boundary condition definition geometry}
        \label{fig:pre_bc}
    \end{subfigure}
    \caption{Geometric preprocessing for the boundary conditions. (a) depicts the predefined shapes used to ensure clean boundary regions in the solid geometry and (b) shows the predefined shapes used to establish consistent boundary conditions on synthetic 3D geometries.}
    \label{fig:pre_geo}
\end{figure}

The synthetic 3D bracket geometries and surface mesh data generated using the marching cubes algorithm are smoothed using the PolyNURBS technique in Altair Inspire. This process converts polygonal surfaces into non-uniform rational b-splines (NURBS), making them solid bodies suitable for FEM simulations. However, this conversion can introduce geometric anomalies, especially at the boundary regions. 
Boundary cleanup geometries (Fig.~\ref{fig:subtract_geo}) around the boundary areas are used to clean up the boundary regions of the solid geometry through geometric operations to mitigate these issues. This approach ensures that the boundary conditions are well-defined and consistent across all samples, which is crucial for the accuracy and reliability of the FEM simulations. The synthetic 3D geometries are non-parametric, making defining boundary conditions based on parametric surfaces challenging. This challenge is addressed by combining the synthesized 3D geometry with predefined boundary condition geometry, as shown in Fig.~\ref{fig:pre_bc}, through geometric operations, resulting in a well-defined FEM model.

\begin{figure} 
\centerline{\includegraphics[width=3.34in]{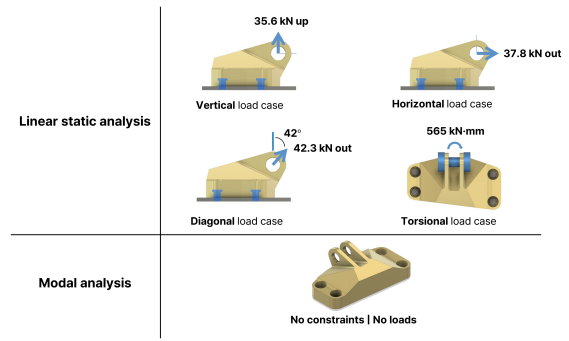}}
\caption{Load conditions for linear static and modal analysis}
\label{fig:load_case}
\end{figure}

Figure~\ref{fig:load_case} illustrates the load conditions used in the simulation, adhering to the structural considerations defined by the GE jet engine bracket CAD challenge. The simulation performs linear static load cases and free–free condition (no constraints or loads) modal analysis. Various loads are applied for the linear static load cases: a vertical load of 35.6 kN applied along the $+z$ axis, a horizontal load of 37.8 kN applied along the $+x$ axis, a diagonal load of 42.3 kN applied 42° from the vertical direction, and a torsional load of 565 kN·mm applied along the $-z$ axis. In these cases, nodes in the bolted interfaces are fixed using RBE2s, and nodes in the loaded interface receive distributed loads via RBE3s.

Modal analysis is also performed under free–free conditions to evaluate the natural frequencies and mode shapes through eigenvalue analysis. This analysis helps in understanding the dynamic behavior of the bracket designs, which can be used to evaluate their reliability and performance under varying operational conditions. 

\subsection{FEA Simulation Results}
The execution of the FEM simulations for each defined load case involves extracting various scalar and field results, which are critical for understanding the structural behavior and performance of the synthetic bracket designs. 

The FEM simulations are executed to gather an extensive range of data. Scalar results, which include body properties, such as mass, volume, and the center of gravity, are extracted. Key performance indicators, such as maximum displacement and maximum von Mises stress values are also recorded as scalar results. These scalar values provide a concise summary of the overall structural performance of each bracket design under the applied load conditions.

However, during the simulation process, a number of cases resulted in failures, totaling 227 instances. These failures were primarily due to convergence errors encountered during the conversion of surface mesh data into NURBS geometry using the PolyNURBS technique, errors in the Boolean operations for boundary condition application, and technical issues such as solver time-out errors and memory allocation failures. These failures were not anticipated during the initial setup, and the absence of quantitative metrics for each failure type reflects the unforeseen nature of these issues. Identifying these failure cases and their underlying causes provides valuable insights that could inform future improvements to the robustness and reliability of the automated simulation pipeline.

\begin{figure} 
\centerline{\includegraphics[width=3.34in]{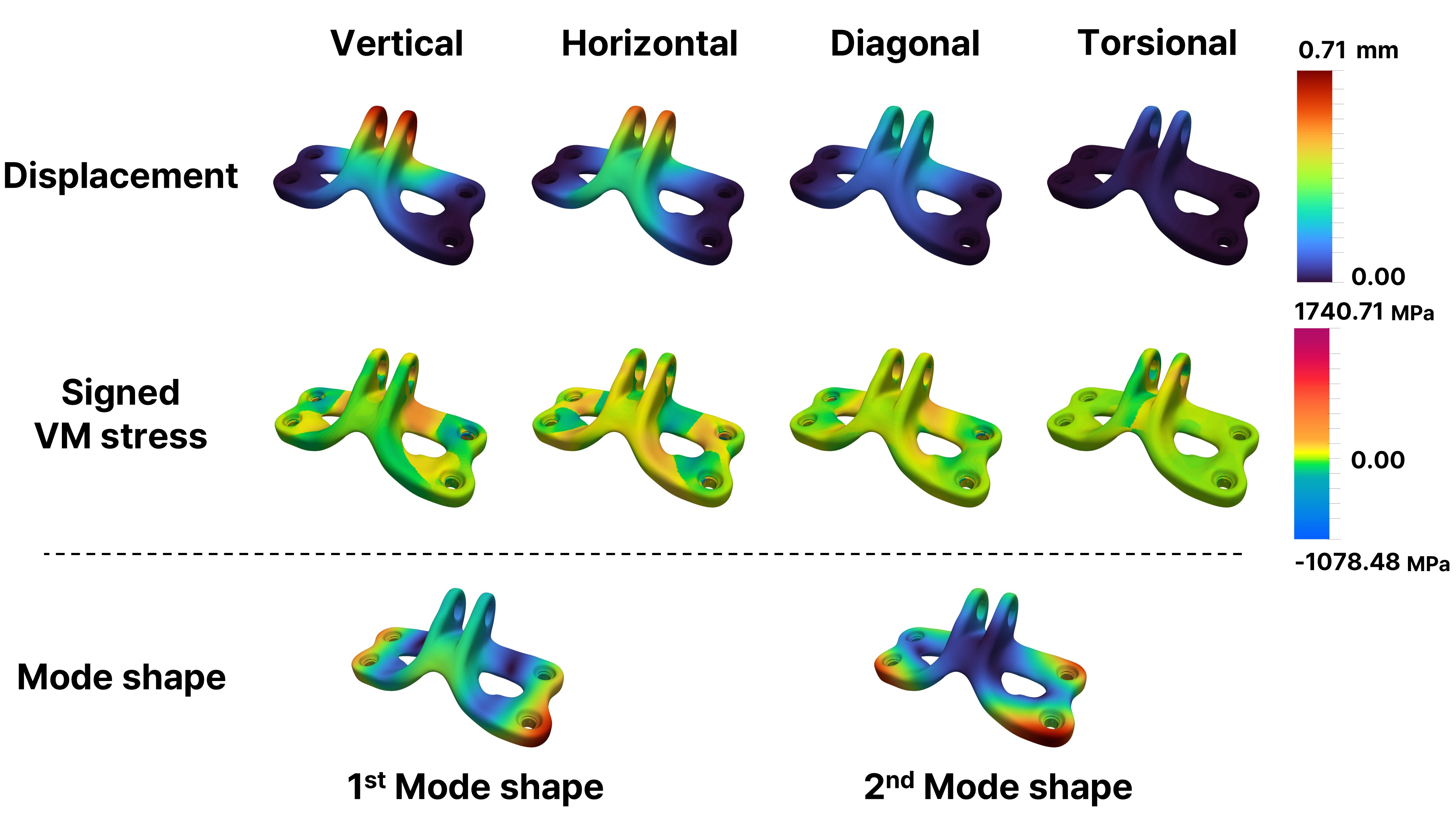}}
\caption{Simulation results. Displacement and signed von Mises stress under vertical, horizontal, diagonal, and torsional loads, along with the first and second normal mode shapes under free–free modal analysis.}
\label{fig:analysis_results}
\end{figure}

Figure~\ref{fig:analysis_results} presents the successful simulation results, offering an in-depth view of the structural response by recording displacement and von Mises stress at every mesh node during linear static analyses. These detailed data demonstrate how different bracket parts react to various loads. Signed von Mises stress is used to indicate the directionality of the stress to provide additional insights, distinguishing between tensile and compressive states within the structure. These directional stress data are crucial for identifying potential failure points and understanding the distribution of stresses throughout the bracket.

In the modal analysis, natural frequencies are extracted as scalar results, while mode shapes are recorded as field results (Fig.~\ref{fig:analysis_results}). These results can be used to evaluate the dynamic behavior of the bracket designs. The natural frequencies measure the bracket's inherent vibrational characteristics, while the mode shapes illustrate how different parts of the bracket deform under these conditions. The analysis explicitly excludes the six rigid body modes, focusing instead on extracting the first and second normal modes.

\begin{figure} 
\centerline{\includegraphics[width=3.34in]{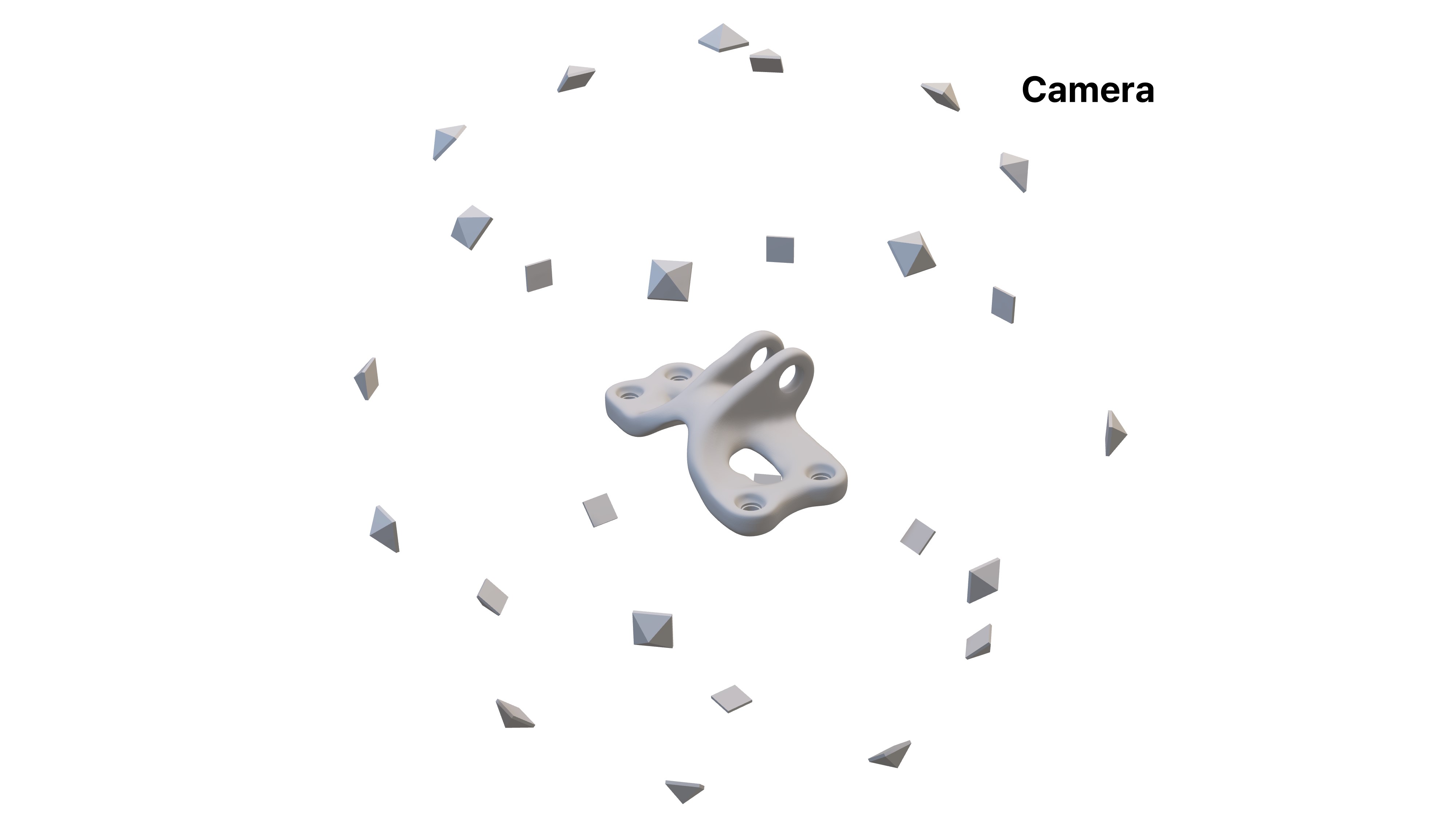}}
\caption{Multi-view images. Bracket objects are surrounded by virtual cameras positioned at 8 azimuth angles and 3 elevation angles, along with additional views from the top and bottom, capturing a total of 26 images for each bracket model for multi-view learning.}
\label{fig:multiview}
\end{figure}

We provide multi-view images that can be applied to multi-view models to contribute to the current interest in the 3D graphics domain. Virtual cameras were positioned at 8 azimuth angles (0°, 45°, 90°, ..., 315°) and 3 elevation angles (45°, 0°, -45°), maintaining a consistent distance from the model center. Additional views were captured from the top and bottom to ensure comprehensive coverage. The images were rendered with consistent lighting settings to ensure uniformity across all views. This process is illustrated in Fig.~\ref{fig:multiview}, which shows the bracket model surrounded by the camera positions used to generate the dataset. These images are an effective alternative to 3D representations, providing a simplified yet comprehensive perspective of the 3D structure. Consequently, the multi-view images can be valuable tools for constructing efficient surrogate models. 

The comprehensive data extracted from these FEM simulations indicate that the DeepJEB dataset is robust and detailed, providing insights into the structural behavior of the synthetic bracket designs. These data can serve as the foundation for developing and validating advanced data-driven models in structural analysis.

\section{Dataset Validation}
\label{Dataset Validation}

The validation of the DeepJEB dataset is essential to ensure its reliability and comprehensiveness as benchmark data for data-driven models. This section details the steps to validate the dataset, including statistical evaluation, outlier removal, geometric quality assessment, and performance space sampling. The applied methodologies are designed to ensure that the DeepJEB dataset meets the high standards for data-driven surrogate modeling.
\subsection{Quality Assurance}
\subsection*{Performance Data Verification}
The performance data verification of the DeepJEB dataset is a critical step in ensuring the dataset's reliability and comprehensiveness for engineering applications. High-quality performance data are crucial for developing accurate data-driven surrogate models, as these data serve as the ground truth labels for input data. Outliers or errors within this data can significantly degrade model performance and reliability, emphasizing the importance of ensuring accurate and clean data.

\begin{figure} 
\centerline{\includegraphics[width=3.34in]{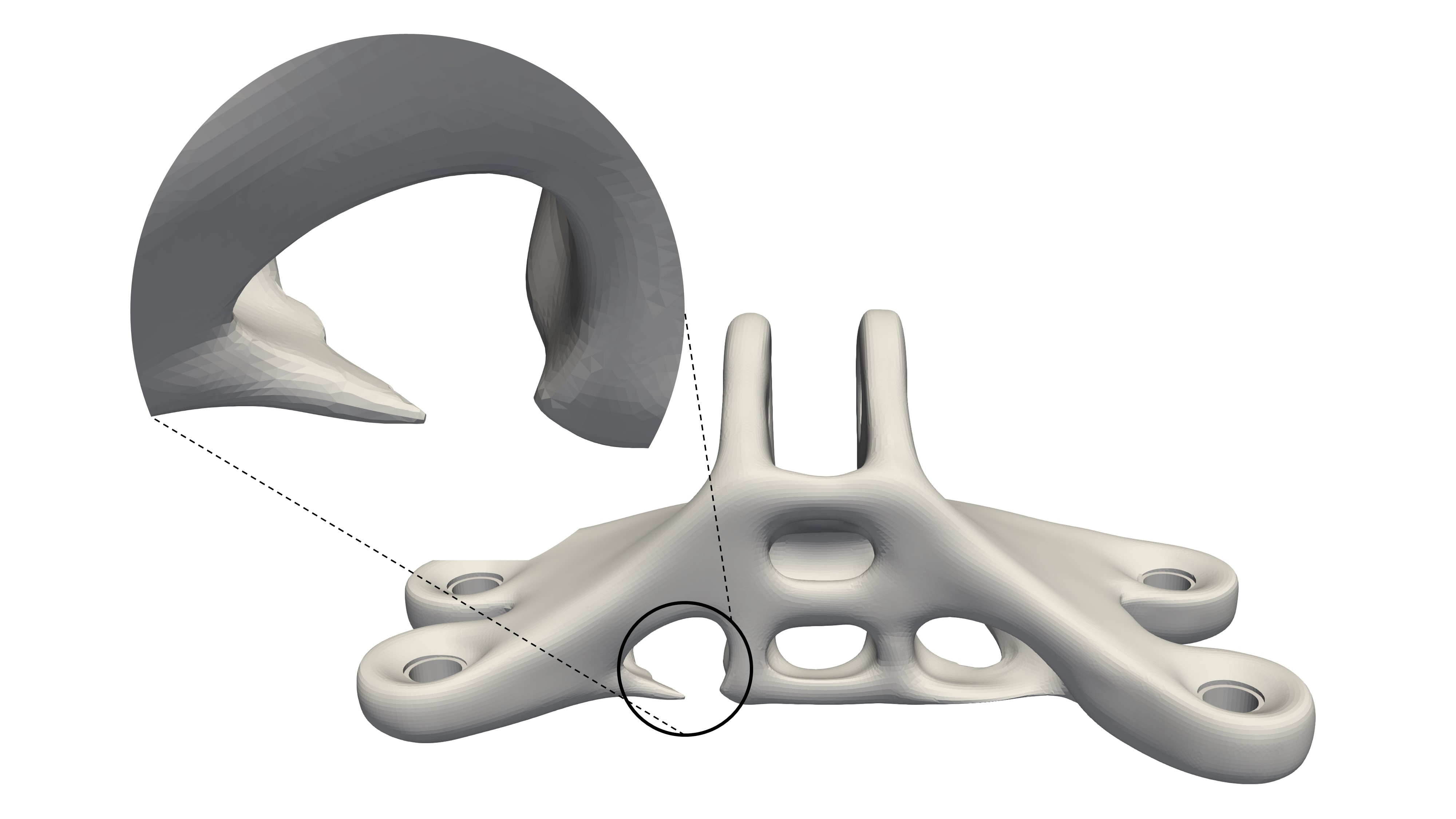}}
\caption{Interpolation artifacts in synthetic data}
\label{fig:interpolation_error}
\end{figure}

Performance data may contain outliers due to issues during the automated simulation process, such as mesh quality problems and geometric operation errors. Figure~\ref{fig:interpolation_error} shows that interpolation artifacts in the synthetic data can result in irregularities in the shape, resulting in unintended discontinuities during numerical simulations. The detection of these errors within automated simulation processes is limited, necessitating the removal of outliers to correct the data. A thorough statistical evaluation of the simulation results is conducted to identify and remove these outliers, ensuring the dataset's reliability.

Balancing geometric diversity while ensuring the reliability of simulation results is essential. This task involves ignoring outliers related to geometric diversity and focusing on outliers in simulation metrics, such as displacement, stress, and natural frequencies. Identifying and removing outliers in these simulation metrics ensures that the dataset remains accurate and reliable while preserving the necessary geometric diversity.

The interquartile range (IQR) method is utilized to detect outliers in performance data. Given the need to preserve geometric diversity, a conservative sensitivity of 3 is used for IQR, corresponding to approximately ±4.72 standard deviations under the assumption of a normal distribution, 0.00024\% of the data fall outside this range. Lower and upper bounds are calculated using the IQR method (Eq.~(\ref{eq:lower_bound}) ~and~(\ref{eq:upper_bound})), to detect and remove outliers in displacement, stress, and natural frequency data. These equations represent the lower and upper bounds for identifying outliers. Samples identified as outliers based on these criteria are removed.

\begin{align}
    \text{IQR} &= Q_3 - Q_1 \notag \\
    \text{Lower Bound} &= Q_1 - 3 \cdot \text{IQR} \tag{3} \label{eq:lower_bound}\\
    \text{Upper Bound} &= Q_3 + 3 \cdot \text{IQR} \tag{4} \label{eq:upper_bound}
\end{align}

\begin{figure}[h!]
    \centering
    \begin{subfigure}[tbh]{3.34in}
        \includegraphics[width=\textwidth]{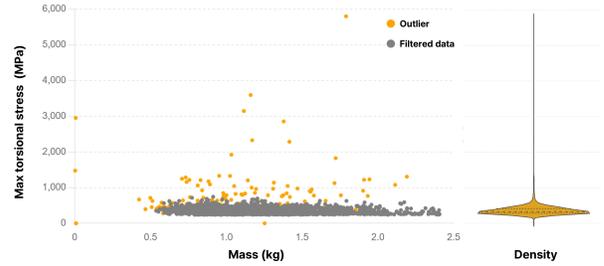}
        \caption{Outlier distribution}
        \label{fig:outliercombined}
    \end{subfigure}
    \begin{subfigure}[tbh]{3.34in}
        \includegraphics[width=\textwidth]{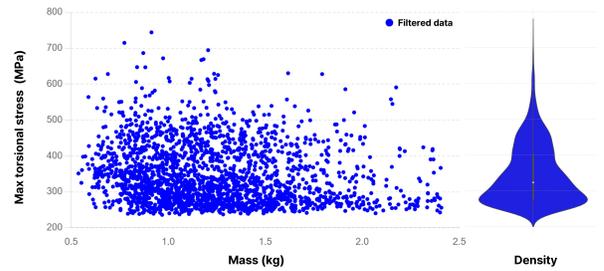}
        \caption{Filtered data distribution}
        \label{fig:filtered}
    \end{subfigure}
    \caption{Distribution of outliers and filtered data. (a) Initial dataset showing outliers affecting the distribution of maximum torsional stress. (b) Filtered dataset demonstrating a more uniform distribution of stress values.}
    \label{fig:outlier}
\end{figure}

After applying the IQR method, 372 samples were identified as outliers and removed from the dataset. This process ensured that the remaining data were of high quality and suitable for reliable surrogate model training. Figure~\ref{fig:outliercombined} illustrates the initial dataset's scatter and violin plots, showing a long tail in the distribution of maximum torsional stress results due to outliers. After filtering, the DeepJEB dataset (Fig.~\ref{fig:filtered}) displayed a uniform distribution around the median values in the scatter and violin plots, indicating a balanced and reliable dataset.

\begin{figure*}[h!]
\centerline{\includegraphics[width=6.50in]{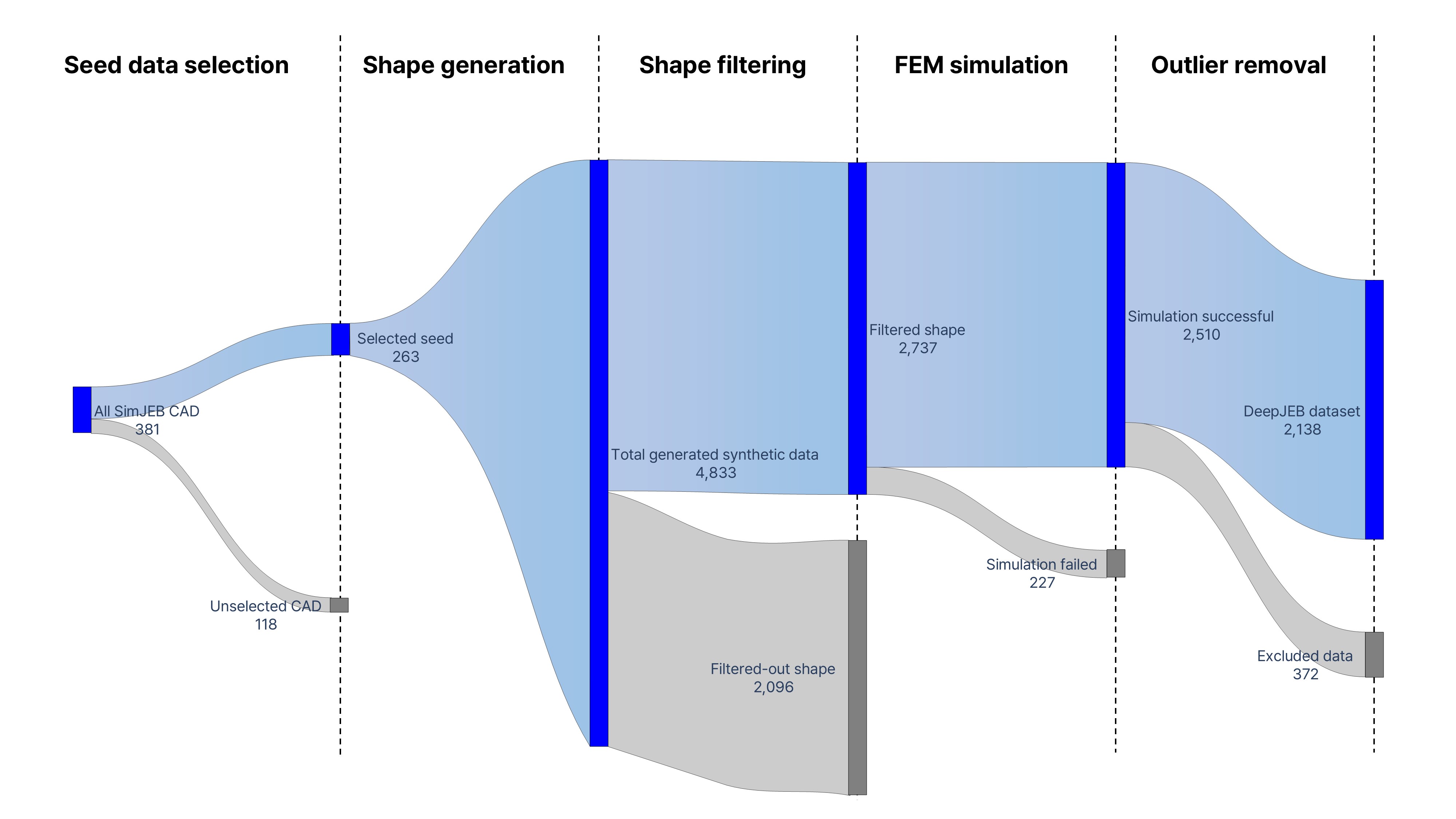}}
\caption{Data generation and filtering workflow. After completing all filtering and validation processes, a total of 2138 DeepJEB data were finalized.}
\label{fig:data_gen_flow}
\end{figure*}

\subsection*{Geometrical Feasibility}
This validation ensures the dataset's integrity and usability, confirming that the final dataset is free from geometric flaws, and the mesh quality is suitable for simulations. Several stages in the data generation process can introduce potential issues, necessitating thorough validation. Non-parametric synthesis with deep generative models can result in irregular shapes, while reconstructing 3D shapes from implicit functions can introduce errors. Smoothing and converting polygonal surfaces to NURBS may cause local geometric distortions, and Boolean operations for boundary conditions can result in complex geometric elements with local inconsistencies. Converting these geometries into volume finite element (FE) meshes can introduce geometric errors despite initial filtering.

The final 3D geometries in the DeepJEB dataset are converted into FE meshes and mapped to simulation results. Evaluation of the quality of these 3D FE meshes is essential to ensure their geometric integrity and suitability for accurate simulations. The evaluation follows industry standards for 3D mesh quality, such as those outlined in \cite{abambres2016finite}. The aspect ratio is a key metric for assessing the quality of 3D mesh elements. An ideal aspect ratio is one, but the aspect ratio must be three or less for stress analysis. In our dataset, only approximately 0.42\% of the mesh elements exceed this aspect ratio threshold, indicating high overall mesh quality. The quality evaluation results show that the DeepJEB dataset consists of 3D meshes with an average of 130,000 elements each. Despite the large number of elements, the low percentage exceeding the aspect ratio threshold demonstrates that most elements meet the quality criteria, maintaining a high standard of geometric quality.

Figure~\ref{fig:data_gen_flow} illustrates the workflow of data creation. After completing all filtering and validation processes, 2138 DeepJEB datasets were finalized.

\subsection*{Limitations}
Despite using second-order elements and more accurate algorithms, FEM-derived performance data can still be potentially inaccurate. A common issue is the inaccuracy of stress results, particularly in stress concentration areas. Although controlling this issue in an automated simulation pipeline is challenging, mesh refinement in these areas can somewhat mitigate the issue. Additionally, even though the filtered samples enhance the reliability and stability of the data, they may not cover all possible designs that could emerge in the context of jet engine bracket interfaces. However, within the defined latent space, the dataset samples are densely populated and suitable for surrogate modeling of similar bracket designs.

\subsection{Dataset Characteristics}
\subsection*{Exploration of the Latent Space}
In this study, data from the baseline dataset served as a starting point for training and generating the DeepJEB dataset. The initial baseline dataset comprised 381 samples, which were reduced to 263 after filtering for consistent boundary conditions. This filtering ensured that only samples with well-defined and consistent boundary conditions were retained, forming a reliable foundation for further data augmentation.

The DeepJEB dataset, augmented based on the 263 selected samples, was evaluated for its geometric diversity to ensure that it adequately represents a wide range of shapes. A $\beta$-variational autoencoder ($\beta$-VAE) was utilized as a dimensionality reduction model to visually assess the shape diversity \cite{higgins2017beta}. This model helps in understanding the distribution and diversity of the dataset by mapping high-dimensional data into a lower-dimensional space, making it easier to analyze and visualize.

In this study, we transformed shapes into SDF grids to evaluate the diversity of the generated data and to conduct performance experiments. While higher-resolution SDF grids can more accurately capture the details of shapes, they also increase data size and model complexity, making it essential to choose an appropriate resolution. To determine a resolution that balances sufficient shape feature representation with manageable memory usage, we converted the DeepJEB data into SDF grids, reconstructed the shapes using the marching cubes algorithm, and then calculated the Chamfer distance (CD). The CD is a metric for measuring the similarity between two point sets by summing the distances from each point in one set to the nearest point in the other set, as defined in Eq.~(\ref{eq:cd}). 
\begin{equation}
\label{eq:cd}\tag{5}
    CD(X,Y)=\sum_{x \in X} \min\limits_{y \in Y} d(x,y) + \sum_{y \in Y} \min\limits_{x \in X} d(x,y)
\end{equation}
The distance between shapes was computed using $d(x,y) = ||x - y||^2$. We calculated the CD between the ground truth data in the dataset and the data reconstructed from SDF using the marching cubes algorithm. As shown in Table~\ref{tab:reso_fidelity}, the CD decreases rapidly up to a resolution of $64^3$, after which the rate of decrease begins to taper off. Based on these observations, we determined that a resolution of $64^3$ is sufficient to capture the essential characteristics of the data while remaining memory-efficient.

The steps for diversity evaluation began with converting the mesh data into SDF grids with a resolution of $64^3$. This conversion provided a standardized representation of the 3D shapes, facilitating subsequent dimensionality reduction processes. The $\beta$-VAE was then used to reduce the dimensionality of the mesh data, aiming to capture the essential features and variations within the dataset.

\begin{table}[h]
\caption{Resolution fidelity check} 
\label{tab:reso_fidelity}
\begin{center}
\begin{tabular}{ccc}
\hline
Resolution &  CD (Mean) & CD (Median)\\
\hline
$16^3$ &  146.9874 & 137.0052   \\
$32^3$ &  19.2286 & 18.5535   \\
$64^3$ &  9.4563 & 8.7014   \\
$128^3$ &  8.7433 & 8.0288   \\
$256^3$ &  8.6607 & 7.9507   \\
\hline
\end{tabular}
\end{center}
\end{table}

\begin{figure}[h!]
\centerline{\includegraphics[width=3.2in]{figures/e8_exp_arch.JPG}}
\caption{3D $\beta$-VAE architecture}
\label{fig:vae_arch}
\end{figure}

DeepJEB and SimJEB datasets were trained using the architecture depicted in Fig.~\ref{fig:vae_arch}. The loss function consisted of a reconstruction loss term and a KL divergence term, as defined in Eq.~(\ref{eq:loss_exp}). 
\begin{equation}
\label{eq:loss_exp}\tag{6}
L_{VAE} = \| p_{\psi}(z_i) - x_i \|_2 + \beta KL(q_{\phi}(z_i|x_i)||p_{\psi}(z))
\end{equation}
The $\beta$-VAE consists of an encoder ($q_{\phi}$) and a decoder ($p_{\psi}$). The loss function $L_{VAE}$ used to train the $\beta$-VAE is composed of two parts: a reconstruction term and a regularization term. The reconstruction term calculates the L2 norm between the output generated by the decoder when given the latent code ($z_i$), and the ground truth data ($x_i$). The regularization term ensures that the distribution of the latent space produced by the encoder when given $x_i$ is similar to the prior distribution $p_{\psi}(z)$. To enhance disentanglement, we set the parameter $\beta$ to 3, which allowed for a clearer separation of the underlying features within the latent space.

\begin{figure}[h!]
\centerline{\includegraphics[width=3.2in]{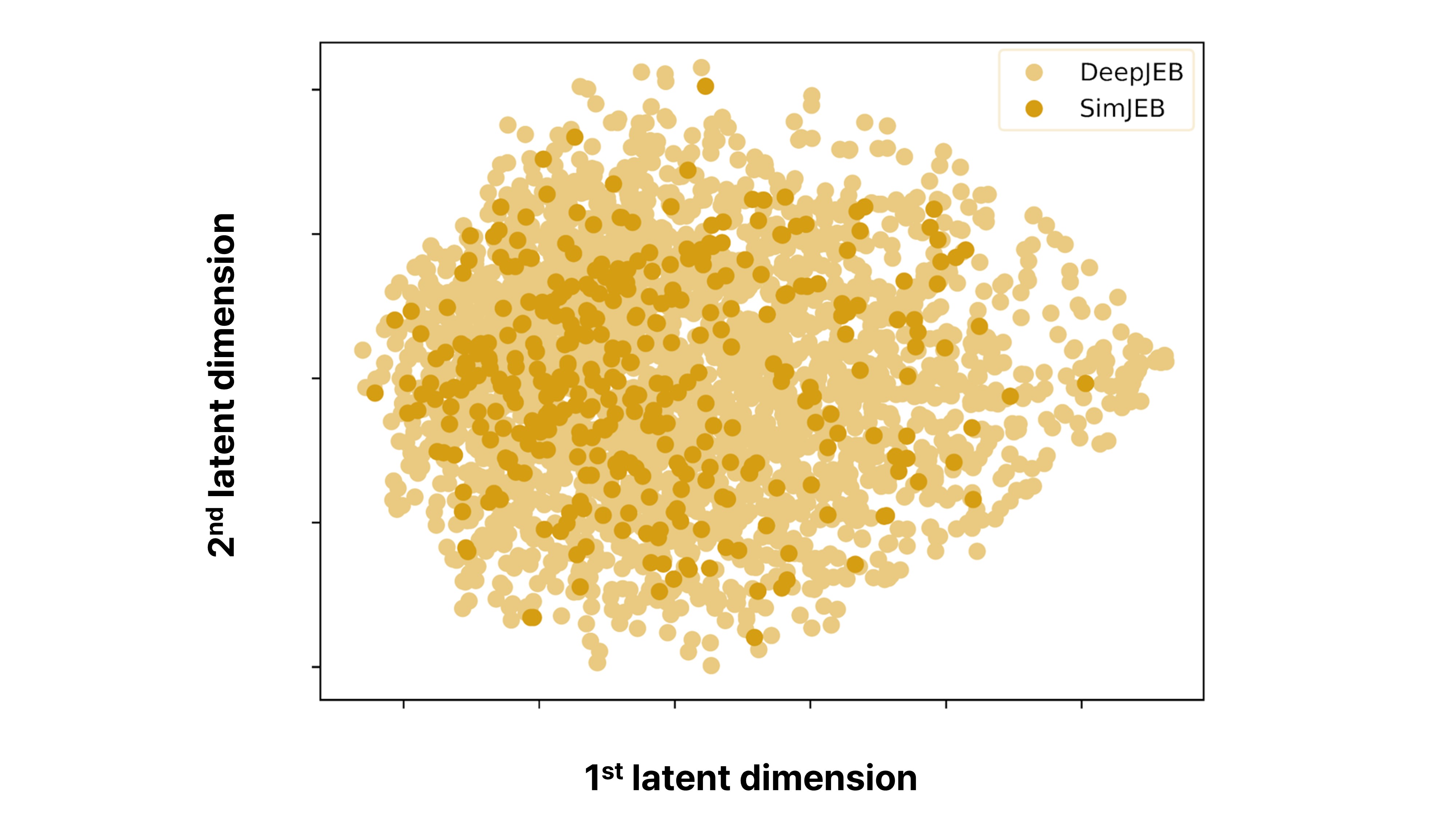}}
\caption{Latent design space. Scatter plot comparing the latent design space of the baseline SimJEB samples and the DeepJEB dataset}
\label{fig:latent_comparison}
\end{figure}

Further dimensionality reduction was performed using principal component analysis (PCA), which enabled visualization of the latent space distribution. As shown in Fig.~\ref{fig:latent_comparison}, scatter plots using the first and second principal components reveal that the latent space of the DeepJEB dataset exhibits a broader and denser distribution compared to the baseline dataset. This suggests that the $\beta$-VAE model was successful in generating a diverse set of latent representations based on the SimJEB subset, potentially reflecting a wide array of geometric features and configurations.

However, it is important to clarify that while the latent space of the DeepJEB dataset shows a wider and denser distribution, this does not necessarily imply an increase in the actual geometric diversity of the dataset compared to SimJEB. The broader distribution in the latent space primarily indicates that the $\beta$-VAE model has enhanced the variability of the latent representations, capturing a range of potential design features that extend beyond the original dataset. This reflects the model's ability to explore and generate a wider array of latent designs, which can be beneficial for various engineering applications requiring robust and flexible datasets.

The expanded latent space distribution of DeepJEB highlights its potential as a comprehensive resource that can better support a wide range of modeling and simulation tasks. The dataset's ability to encompass more diverse latent representations makes it well-suited for applications where a rich variety of design configurations is crucial. While this broader distribution does not directly translate to increased geometric diversity, the enriched latent space allows for a more extensive exploration of design possibilities, thereby offering a valuable tool for further research and development.

It is also essential to consider the limitations imposed by the data selection process. In constructing the DeepJEB dataset, we employed a seed data selection process that excluded structures with thin features or cavities, which could generate unstable noise during the data synthesis. Although this approach improved the dataset's overall quality and stability, it may have reduced the range of geometric configurations represented, potentially limiting the dataset's ability to fully capture all possible geometric variations from the original SimJEB dataset. Thus, while DeepJEB provides a valuable extension in the latent representation space, these limitations should be considered when applying it to various engineering and modeling tasks.

In conclusion, the results demonstrate that DeepJEB, despite being derived from a subset of SimJEB, offers a valuable extension in the latent representation space, enhancing exploration capabilities for design optimization and analysis. The dataset's utility lies in its potential to serve as a versatile benchmark for various surrogate modeling and structural analysis applications, ensuring robust performance across different scenarios.

\subsection*{Testset Composition}
\label{testset}

\begin{figure*}[h!]
\centerline{\includegraphics[width=6.5in]{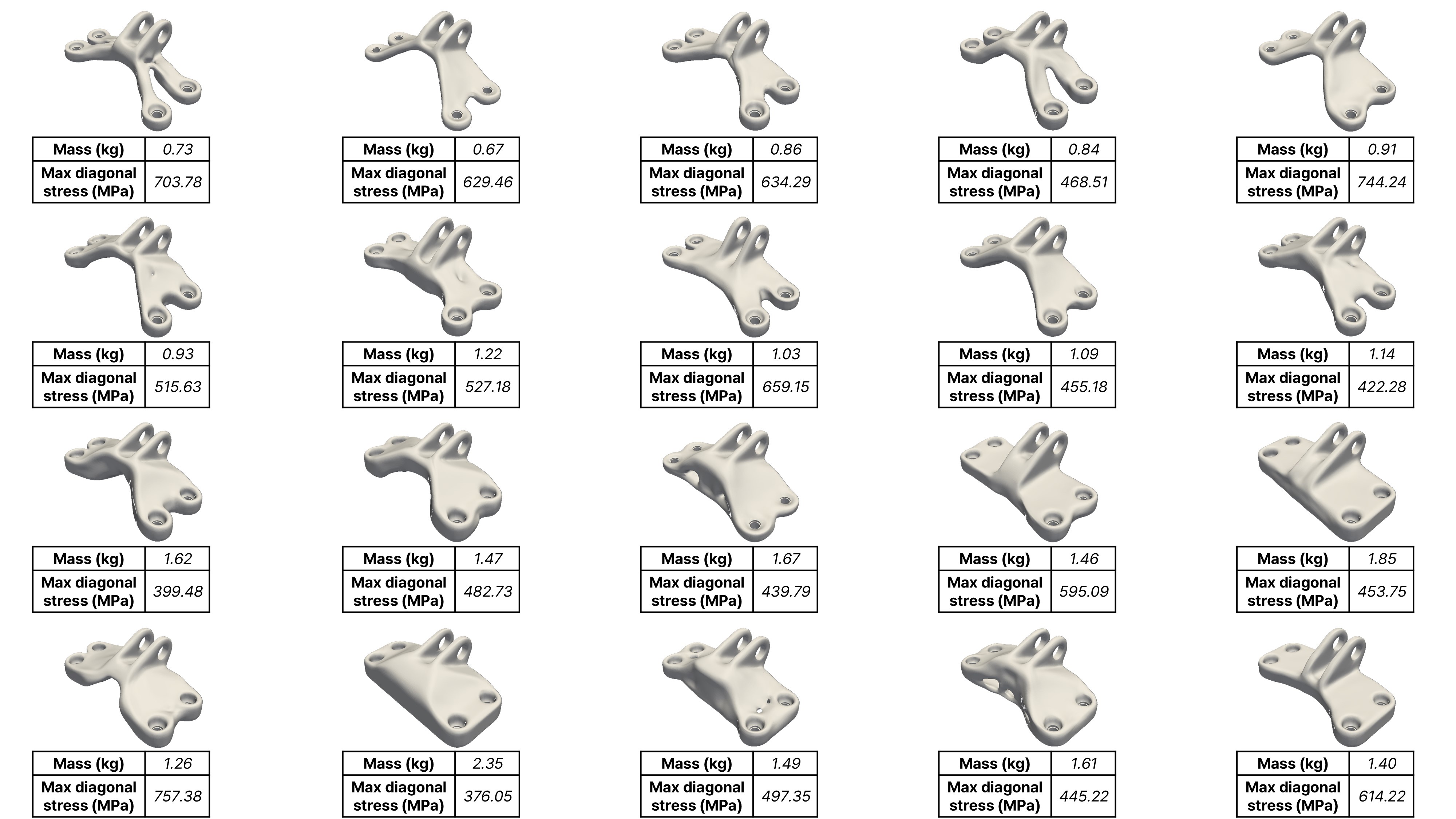}}
\caption{Cluster representative samples. Sample geometries representing each of the 20 clusters formed using k-means clustering}
\label{fig:cluster_samples}
\end{figure*}

Given the diverse geometric variations in the DeepJEB dataset, unsupervised clustering techniques are utilized to effectively classify and sample the latent space. These algorithms group unlabeled data samples based on their inherent shape similarities, ensuring a comprehensive latent space representation. Key features are extracted to manage the complexity of geometric data, and dimensionality is reduced using $\beta$-VAE. This method preserves critical information while simplifying the data for clustering. The process begins by converting 3D shape data to $64^3$-resolution SDF grids and then using $\beta$-VAE for dimensionality reduction, which maintains essential geometric features.

The reduced latent vectors are classified using k-means clustering. silhouette score and the elbow method are applied to determine the optimal number of clusters, resulting in 20. The steps involve determining the optimal number of clusters (k=20) using silhouette scores and the elbow method, followed by applying k-means to form 20 distinct clusters based on geometric features. Figure~\ref{fig:cluster_samples} presents sample geometries representing each of the 20 clusters and their corresponding performance metrics. Each cluster represents a unique segment of the latent space, and uniform sampling within these clusters ensures that all possible design variations are included in the test set. This process involves forming 20 clusters representing different design variations and using uniform sampling within each cluster to ensure comprehensive coverage of the design space.

\begin{figure*}[h!]
    \centering
    \begin{subfigure}[tbh]{6.60in}
        \centerline{\includegraphics[width=6.60in]{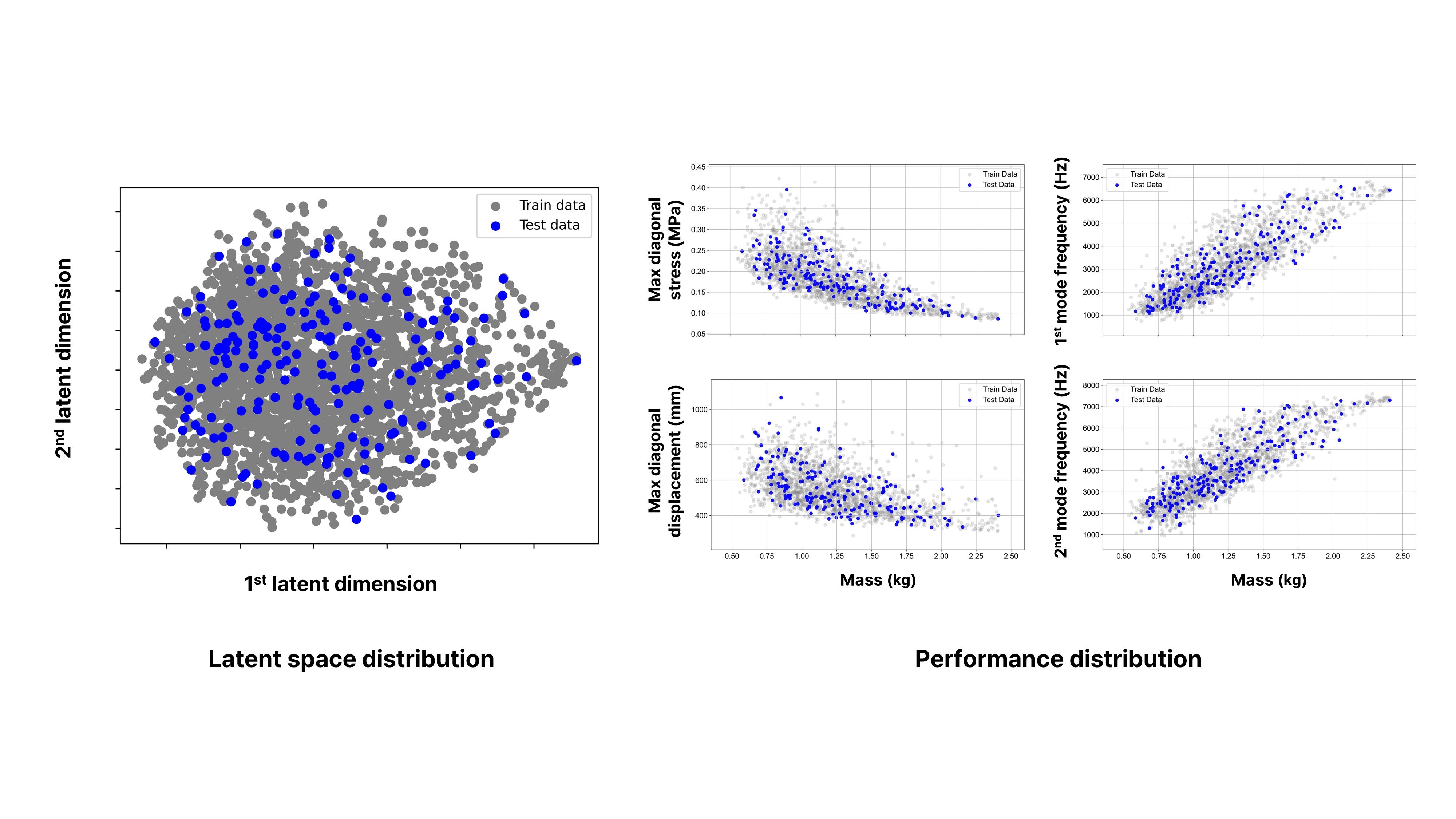}}
        \caption{Latent space sampled test set distribution}
        \label{fig:latent_sample_distribution}
    \end{subfigure}
    \begin{subfigure}[tbh]{6.60in}
        \centerline{\includegraphics[width=6.60in]{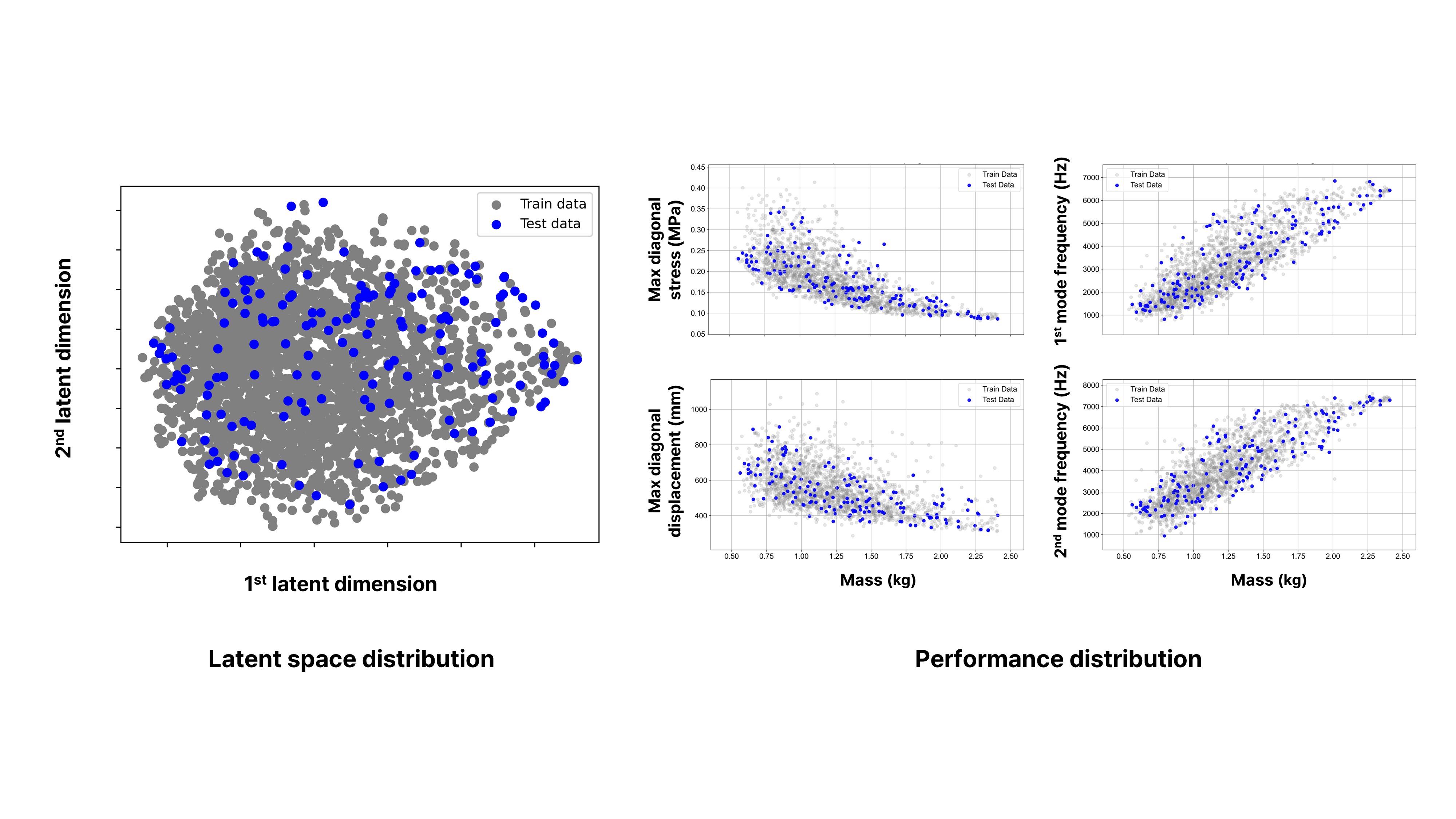}}
        \caption{Performance space sampled test set distribution }
        \label{fig:performnace_sample_distribution}
    \end{subfigure}
    \caption{Test set distribution in the latent and performance spaces. These plots demonstrate that the test sets are evenly distributed across both the latent and the performance spaces. The performance space includes metrics, such as maximum diagonal stress, maximum diagonal displacement, and the first and second mode frequencies.}
    \label{fig:sample_distribution}
\end{figure*}

Unsupervised clustering and uniform sampling ensure that the test set of the DeepJEB dataset accurately represents the entire latent space. Figure~\ref{fig:latent_sample_distribution} shows the latent space and performance distribution of the test set. This representation guarantees that all possible design variations are included, enhancing the reliability and applicability of surrogate models trained on this dataset.

Performance labels represent specific performance metrics for each design, such as maximum displacement, minimum–maximum stress, and natural frequency. These multi-dimensional performance data are reduced using PCA to manage the complexity and ensure a comprehensive representation. PCA captures the most significant variations in the performance data, simplifying it to principal components. The steps include identifying performance metrics, such as displacement, stress, and natural frequency, and applying PCA to reduce the dimensionality of the performance data, focusing on key components that capture the most variance.

Uniform sampling is applied within the reduced performance distribution to ensure that the performance space is comprehensively represented. This method ensures that the test set reflects all possible performance ranges, providing a robust basis for evaluating model performance. The process involves using PCA to create a reduced-dimensional performance distribution and applying uniform sampling to select representative performance combinations within this space. 

The generated test set, derived from uniform sampling in the performance space, accurately represents the performance distribution (Fig.~\ref{fig:performnace_sample_distribution}). This test evaluates various performance scenarios, ensuring the model learns from diverse performance data and enhancing its generalization capabilities.

The test sets are provided in JSON format, including metadata for each 3D model's name. This format enhances accessibility and compatibility, making it easy to use across various systems and software. 

\section{Case Study: Surrogate modeling of Structural Performance}
\label{Case Study}

In this section, we delve into the practical application of the DeepJEB dataset for developing surrogate models to predict structural performance. We illustrate the dataset's utility through a comparative analysis with the baseline dataset, highlight the benefits of using synthetic data, and validate model performance using a uniformly sampled test set in latent and performance spaces.

\subsection{Surrogate Model Construction}

\begin{figure}[h!]
\centerline{\includegraphics[width=3.0in]{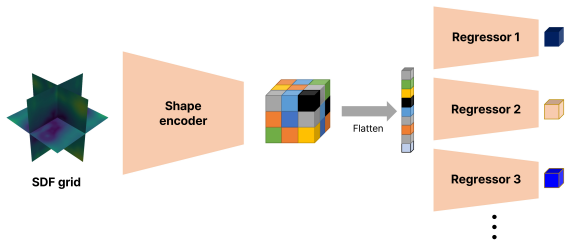}}
\caption{Surrogate model architecture}
\label{fig:eval_arch}
\end{figure}

The surrogate model consists of an encoder that compresses the shape data and a regressor that predicts performance metrics based on the compressed shape features. The model's architecture was designed to assess the performance diversity and validity of the DeepJEB dataset by training it under the same conditions as those used for the baseline dataset. As shown in Fig.~\ref{fig:eval_arch}, the surrogate model comprises a shape-encoding encoder and a set of regressors that predict individual performances. To evaluate uncertainty, we implemented a Deep Ensemble model \cite{lakshminarayanan2017simple}. The Ensemble model comprises five individual models, with the detailed architecture illustrated in Fig.~\ref{fig:eval_arch}. Each model consists of one Encoder and a number of regressors corresponding to the number of labels being predicted. Each regressor produces two outputs: the predicted value ($\mu$) and $\sigma^2$ for uncertainty quantification. The loss term used in this study is defined in Eq.~(\ref{eqn:loss_eval}).
\begin{equation}
\label{eqn:loss_eval}\tag{7}
Loss_{NLL} = \frac{log(\sigma^2_\alpha(x))}{2} + \frac{(y-\mu_\alpha(x))^2}{2\sigma^2_\alpha(x)}+ 1
\end{equation}
Specifically, our study utilizes four regressors, each dedicated to predicting one of the four displacements. The shape encoder receives the $64^3$-resolution SDFs of the preprocessed individual shapes in a grid format and processes them through 3D convolutional layers for learning. The number of regressors corresponds to the number of performances to be predicted, all of which take latent codes compressed by the shape encoder as input. The regressors utilize a simple multi-layer perceptron consisting of linear layers and rectified linear unit functions. The model is trained to predict vertical, horizontal, diagonal, and torsional maximum displacements using these labels to evaluate its accuracy and reliability.

\begin{figure} 
\centerline{\includegraphics[width=3in]{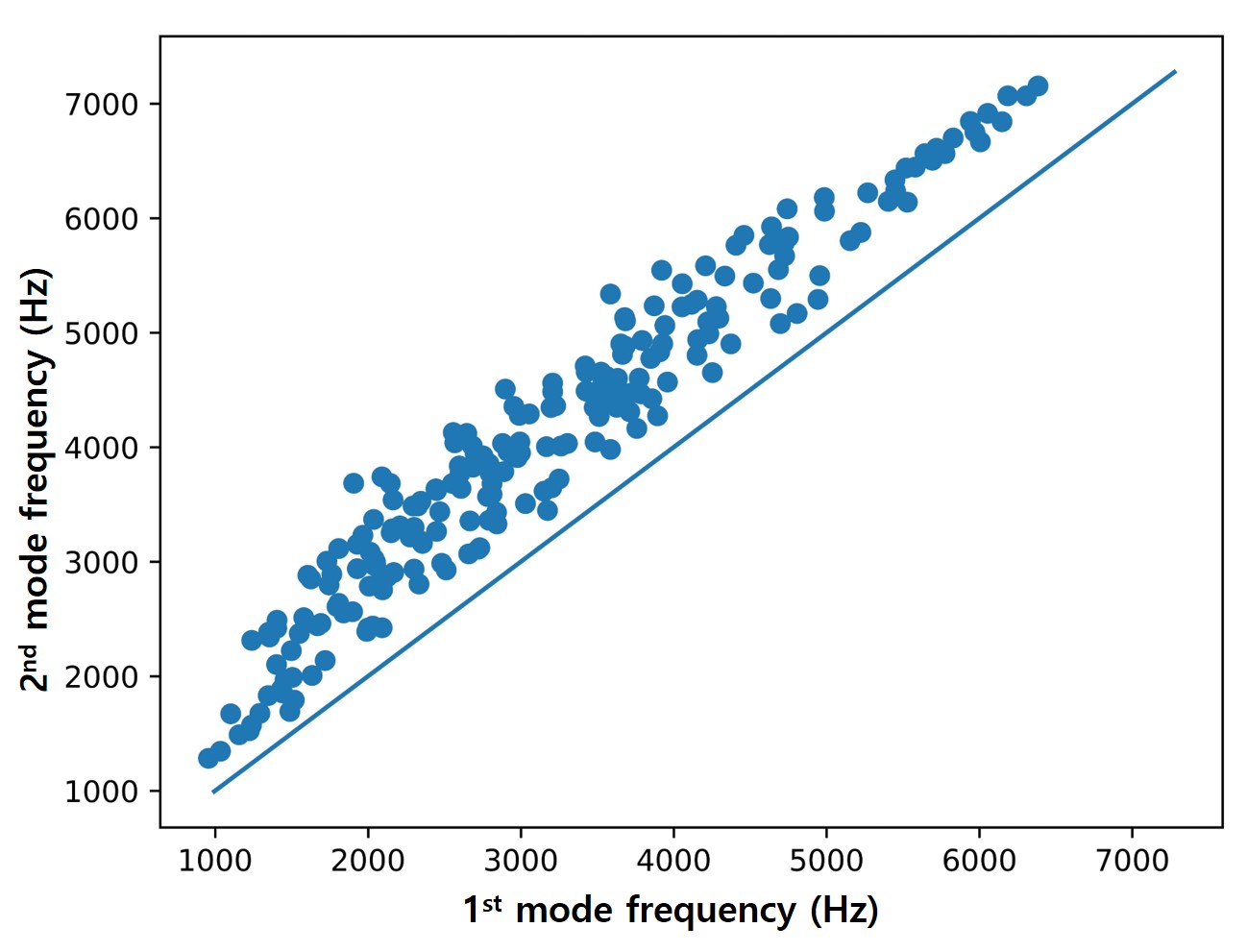}}
\caption{Comparison of the first and second mode frequency}
\label{fig:deepjeb_freq_comparison}
\end{figure}

\begin{figure} 
\centerline{\includegraphics[width=3.34in]{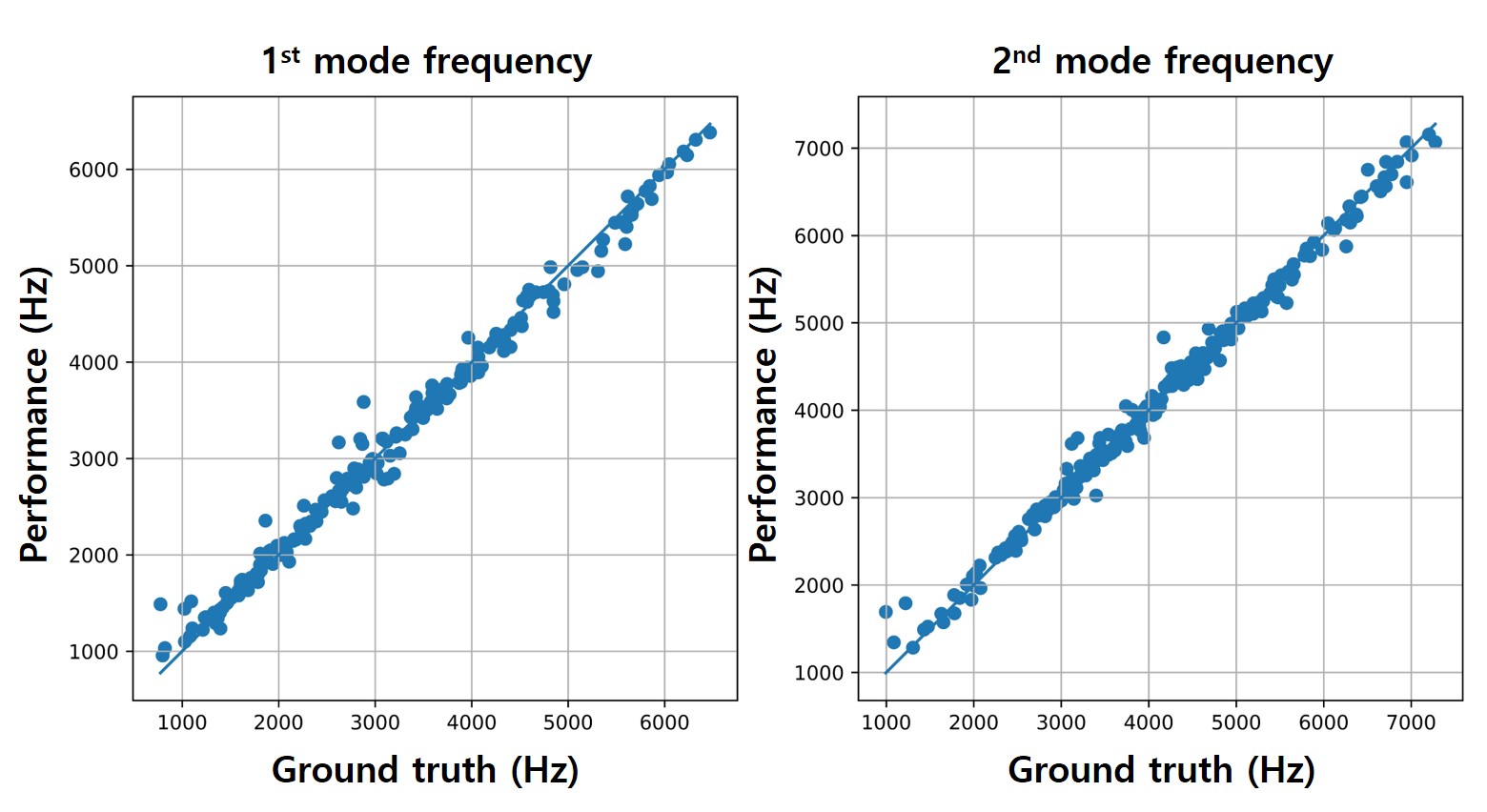}}
\caption{Frequency scatter plot of the trained DeepJEB surrogate model on the test dataset, comparing the predicted first and second mode frequencies to the ground truth.}
\label{fig:deepjeb_scatter_freq}
\end{figure}

Furthermore, we evaluated the surrogate model's performance in predicting the first and second mode frequencies provided by DeepJEB. The characteristic that the second mode frequency is always more significant than the first mode frequency was used to validate the model's understanding of domain-specific information. Key observations included the surrogate model's ability to predict first and second mode frequencies and the reflection of domain-specific characteristics in the model's predictions. We conducted experiments with DeepJEB, dividing the dataset into training, validation, and test sets in an 8:1:1 ratio. DeepJEB demonstrated high accuracy, successfully learning the overall trends (Fig.~\ref{fig:deepjeb_scatter_freq}). Table~\ref{tab:mode_train} shows quantitative confirmation of this high accuracy. Furthermore, DeepJEB effectively learned the domain-specific knowledge that the first mode frequency should be lower than the second mode frequency (Fig.~\ref{fig:deepjeb_freq_comparison}). This mechanism validated the model's capability to incorporate and reflect critical domain-specific characteristics in its predictions.

\begin{table}[h]
\caption{Test results of mode frequency surrogate model} 
\label{tab:mode_train}
\begin{center}
\begin{tabular}{ccc}
\hline
\multirow{4}{*}{\begin{tabular}[c]{@{}c@{}}First mode\\ frequency\end{tabular}} &  R²  & 0.9882\\
&  MAE & 97.64 \\
&  MSE & 2.188 $\times 10^{-4}$   \\
&  MAPE (\%) & 4.142 \\
\hline
\multirow{4}{*}{\begin{tabular}[c]{@{}c@{}}Second mode\\ frequency\end{tabular}} &  R²  & 0.9902 \\
&  MAE & 92.55   \\
&  MSE & 1.965 $\times 10^{-4}$\\
&  MAPE (\%) & 2.936 \\
\hline
\end{tabular}
\end{center}
\end{table}
We compared the performance of surrogate models trained on DeepJEB data with those trained on baseline data to demonstrate the advantages of using synthetic data. This comparison highlighted improvements in prediction accuracy and reliability attributable to the enhanced diversity and detail in the DeepJEB dataset. The comparison showed that surrogate models trained on DeepJEB data exhibited improved prediction accuracy and reliability compared with those trained on the baseline data.

\begin{figure*} [h!]
    \centering
    \begin{subfigure}[tbh]{5.50in}
        \includegraphics[width=\textwidth]{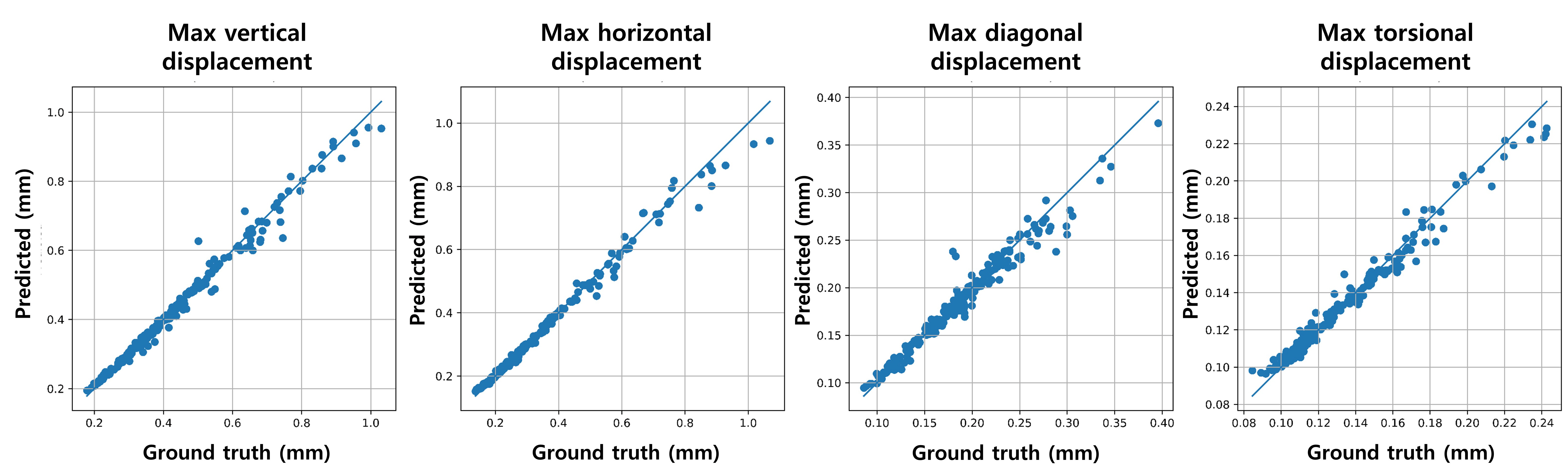}
        \caption{Displacement scatter plot for design diversity in DeepJEB}
        \label{fig:deepjeb_scatter_x}
    \end{subfigure}
    \begin{subfigure}[tbh]{5.50in}
        \includegraphics[width=\textwidth]{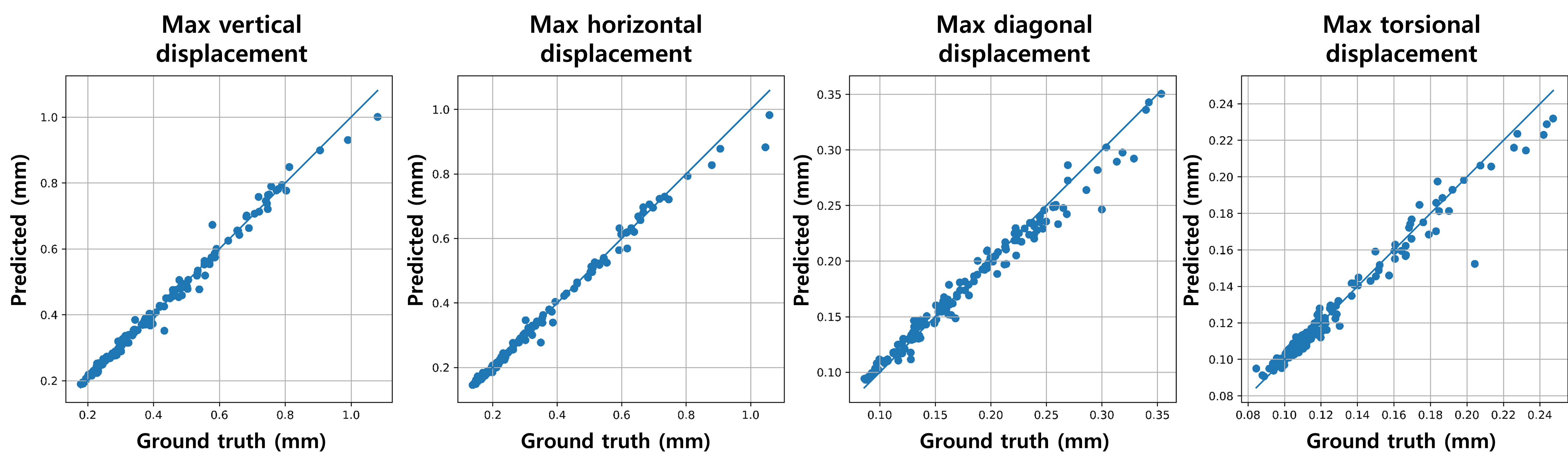}
        \caption{Displacement scatter plot for performance diversity in DeepJEB}
        \label{fig:deepjeb_scatter_y}
    \end{subfigure}
    \begin{subfigure}[tbh]{5.50in}
        \includegraphics[width=\textwidth]{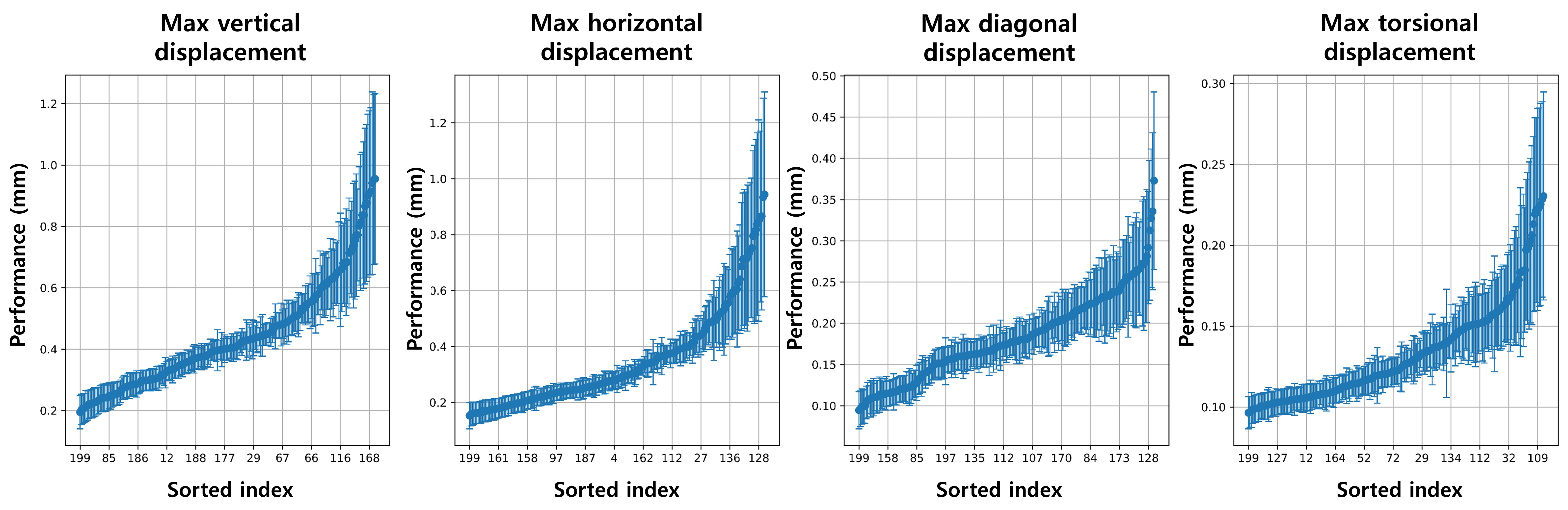}
        \caption{Uncertainty graph for design diversity in DeepJEB displacement test}
        \label{fig:deepjeb_uq_disp_x}
    \end{subfigure}
    \begin{subfigure}[tbh]{5.50in}
        \includegraphics[width=\textwidth]{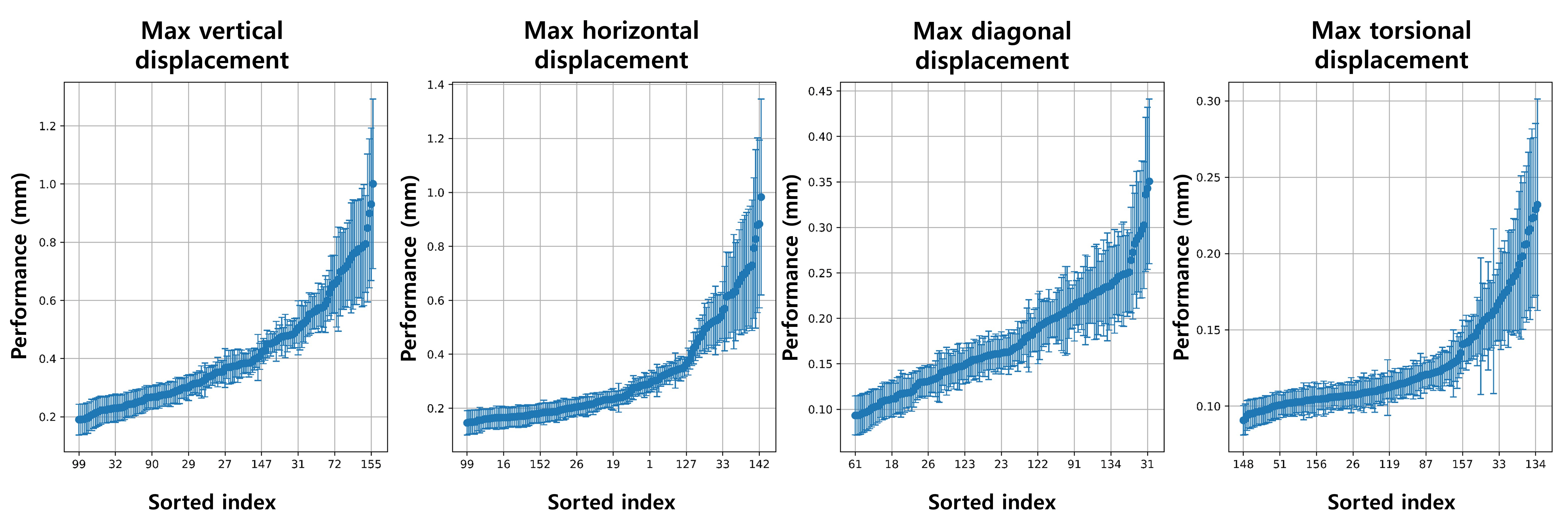}
        \caption{Uncertainty graph for performance diversity in DeepJEB displacement test}
        \label{fig:deepjeb_uq_disp_y}
    \end{subfigure}
    \caption{DeepJEB test results}
    \label{fig:deepjeb_disp_test}
\end{figure*}

\begin{figure*} [h!]
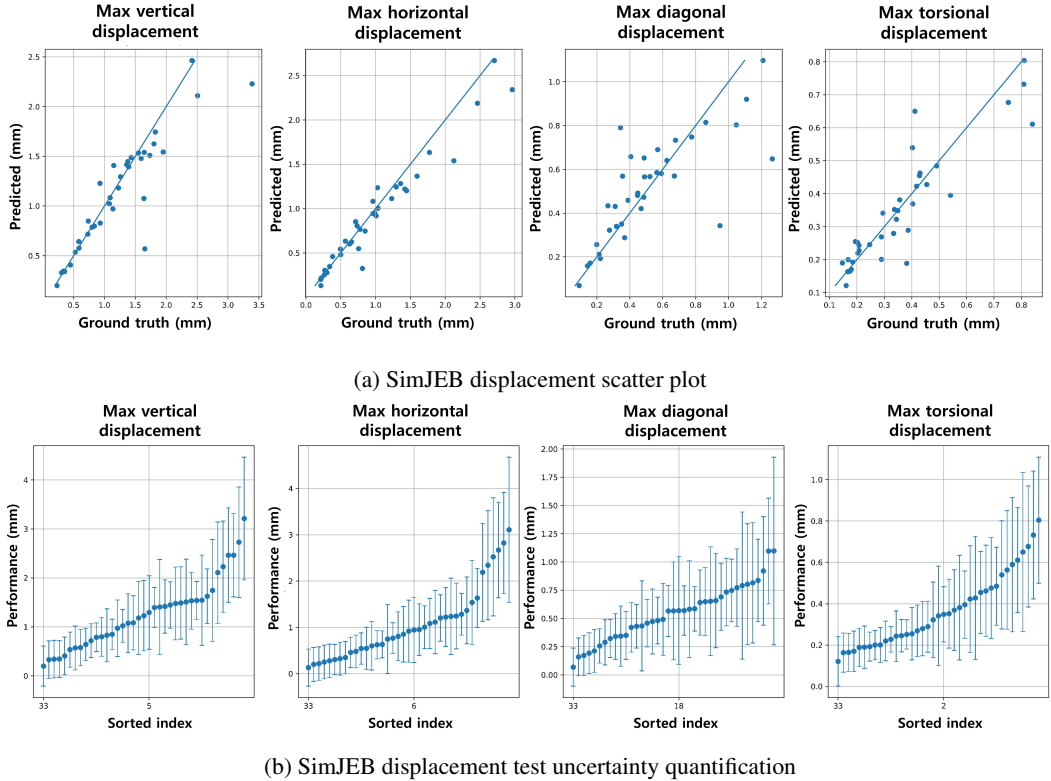

    \centering
    \begin{subfigure}[tbh]{5.50in}
        \includegraphics[width=\textwidth]{figures/e6_sim_scat.JPG}
        \caption{SimJEB displacement scatter plot}
        \label{fig:simjeb_scatter_disp}
    \end{subfigure}
    \begin{subfigure}[tbh]{5.50in}
        \includegraphics[width=\textwidth]{figures/e6_sim_uq.JPG}
        \caption{SimJEB displacement test uncertainty quantification}
        \label{fig:simjeb_uq_disp}
    \end{subfigure}
    \caption{SimJEB test results}
    \label{fig:simjeb_disp_test}
\end{figure*}

Furthermore, we divided the DeepJEB dataset into training, validation, and test sets in an 8:1:1 ratio, following the same experimental protocol as the previous frequency experiments. In this case, for DeepJEB, we used two uniformly sampled test datasets based on the distributions of shapes and performances, as outlined in Section~\ref{testset}. In contrast, SimJEB split the training and test datasets using random sampling. The results of these experiments are as follows: the performance metrics for the experimental datasets demonstrate a significant improvement compared with the baseline data across all performance indicators (Table~\ref{tab:disp_train}--\ref{tab:comp_btw_sim_deep}). The scatter plots in Figs.~\ref{fig:deepjeb_scatter_x} and \ref{fig:deepjeb_scatter_y} illustrate that all test sets are classified according to the shape and performance distributions exhibit strong trend adherence. The uncertainty quantification graph in Figs.~\ref{fig:deepjeb_uq_disp_x} and \ref{fig:deepjeb_uq_disp_y} reveals that the regions with high data sample density, which are known a priori, show reduced uncertainty. Overall, the dataset's high accuracy and reliability are evident when comparing the scatter plot and uncertainty quantification graph in Fig.~\ref{fig:simjeb_disp_test} to those of the SimJEB data. These observations confirm that the surrogate models trained on the DeepJEB dataset provide precise predictions and maintain robustness, as evidenced by the quantified uncertainties. The models effectively capture complex relationships and demonstrate enhanced predictive capabilities and reliability in practical applications. To quantify the performance improvement caused by using a uniform dataset, a DeepJEB sub-dataset of the same size as SimJEB was created, and a predictive model was trained. For this purpose, 381 data points were randomly sampled. The results are shown in Table \ref{tab:comp_btw_sim_deep}. Despite the smaller dataset, the model demonstrated high performance on both training and test sets.

\begin{table*}[t]
\caption{Training results of displacement surrogate model}
\label{tab:disp_train}
\addtocounter{table}{-1}
\begin{adjustbox}{minipage=0.85\textwidth, center}
\begin{tabular}{cccccc}
\hline
\multirow{2}{*}{\begin{tabular}[c]{@{}c@{}}Engineering\\ label\end{tabular}} & \multirow{2}{*}{Metrics}  & \multicolumn{2}{c}{\begin{tabular}[c]{@{}c@{}}DeepJEB\\ (shape uniform)\end{tabular}} & \multicolumn{2}{c}{\begin{tabular}[c]{@{}c@{}}DeepJEB\\ (performance uniform)\end{tabular}} \\ & & Train & Test & Train & Test \\
\hline
\multirow{4}{*}{\begin{tabular}[c]{@{}c@{}}Max vertical\\ displacement\end{tabular}}   & R²  &0.9919& 0.9858& \textbf{0.9952}& \textbf{0.9901}  \\
& MAE &9.335$\times 10^{-3}$ & 1.229$\times 10^{-2}$&\textbf{7.009}$\mathbf{\times 10^{-3}}$ & \textbf{1.205}$\mathbf{\times 10^{-2}}$ \\
& MSE &2.343$\times 10^{-4}$ & 4.585$\times 10^{-4}$&\textbf{1.366}$\mathbf{\times 10^{-4}}$ & \textbf{3.450}$\mathbf{\times 10^{-4}}$ \\
& MAPE (\%) &2.021 & \textbf{2.649} &\textbf{1.565} & 3.157 \\
\hline
\multirow{4}{*}{\begin{tabular}[c]{@{}c@{}}Max horizontal\\ displacement\end{tabular}} & R²  &0.9916 & 0.9882&\textbf{0.9927}& \textbf{0.9901}   \\
& MAE &8.541$\times 10^{-3}$ & 1.068$\times 10^{-2}$&\textbf{6.596}$\mathbf{\times 10^{-3}}$ & \textbf{9.568}$\mathbf{\times 10^{-3}}$ \\
& MSE &2.407$\times 10^{-4}$ &4.082$\times 10^{-4}$&\textbf{2.091}$\mathbf{\times 10^{-5}}$ & \textbf{3.744}$\mathbf{\times 10^{-4}}$ \\
& MAPE (\%) &2.307 & \textbf{2.807}&\textbf{1.657} & 2.937 \\
\hline
\multirow{4}{*}{\begin{tabular}[c]{@{}c@{}}Max diagonal\\ displacement\end{tabular}}   & R²  &\textbf{0.9881} & 0.9597&0.9876& \textbf{0.9738}   \\
& MAE &3.951$\times 10^{-3}$ & 7.015$\times 10^{-3}$&\textbf{3.913}$\mathbf{\times 10^{-3}}$ & \textbf{6.792}$\mathbf{\times 10^{-3}}$ \\
& MSE &\textbf{4.146}$\mathbf{\times 10^{-5}}$ & 1.235$\times 10^{-4}$&4.240$\times 10^{-5}$ & \textbf{9.775}$\mathbf{\times 10^{-5}}$ \\
& MAPE (\%) &2.125 & \textbf{3.651} &\textbf{2.024} & 3.956 \\
\hline
\multirow{4}{*}{\begin{tabular}[c]{@{}c@{}}Max torsional\\ displacment\end{tabular}}   & R²  &0.9881& \textbf{0.9765}&\textbf{0.9895}& 0.9674 \\
& MAE &2.507$\times 10^{-3}$ & \textbf{3.559}$\mathbf{\times 10^{-3}}$&\textbf{1.991}$\mathbf{\times 10^{-3}}$ & 3.677$\times 10^{-3}$ \\
& MSE &1.311$\times 10^{-5}$ & \textbf{2.580}$\mathbf{\times 10^{-5}}$&\textbf{1.143}$\mathbf{\times 10^{-5}}$ & 4.042$\times 10^{-5}$ \\
& MAPE (\%) &1.790 & \textbf{2.604} &\textbf{1.345} & 2.625 \\
\hline
\end{tabular}
\end{adjustbox}
\end{table*}

\begin{table*}[t]
    \addtocounter{table}{1}
    \caption{Comparison between SimJEB and DeepJEB}
    \label{tab:comp_btw_sim_deep}
    \begin{adjustbox}{minipage=0.85\textwidth, center}
    \begin{tabular}{cccccccc}
    \hline
    \multirow{2}{*}{\begin{tabular}[c]{@{}c@{}}Engineering\\ label\end{tabular}} & \multirow{2}{*}{Metrics}  & \multicolumn{2}{c}{\begin{tabular}[c]{@{}c@{}}DeepJEB\\ (same size as SimJEB)\end{tabular}} &  \multicolumn{2}{c}{SimJEB}  \\ & & Train & Test & Train & Test \\
    \hline
    \multirow{4}{*}{\begin{tabular}[c]{@{}c@{}}Max vertical\\ displacement\end{tabular}}   & R²  & \textbf{0.9875}& \textbf{0.9886}& 0.9693 & 0.8049  \\
    & MAE &\textbf{1.225}$\mathbf{\times 10^{-2}}$ & \textbf{1.204}$\mathbf{\times 10^{-2}}$& 8.557$\times 10^{-2}$ & 1.606$\times 10^{-1}$  \\
    & MSE &\textbf{3.860}$\mathbf{\times 10^{-4}}$ & \textbf{2.913}$\mathbf{\times 10^{-4}}$& 1.616$\times 10^{-2}$ & 9.485$\times 10^{-2}$ \\
    & MAPE (\%) & \textbf{2.800} &\textbf{3.383} & 9.085 & 14.906 \\
    \hline
    \multirow{4}{*}{\begin{tabular}[c]{@{}c@{}}Max horizontal\\ displacement\end{tabular}} & R²  &\textbf{0.9896}& \textbf{0.9892}& 0.9800 & 0.8554   \\
    & MAE &\textbf{1.184}$\mathbf{\times 10^{-2}}$ & \textbf{1.285}$\mathbf{\times 10^{-2}}$& 4.848$\times 10^{-2}$ & 1.676$\times 10^{-1}$  \\
    & MSE &\textbf{3.152}$\mathbf{\times 10^{-4}}$ & \textbf{2.469}$\mathbf{\times 10^{-4}}$& 1.238$\times 10^{-2}$ & 1.114$\times 10^{-1}$ \\
    & MAPE (\%) & \textbf{3.716}&\textbf{5.087} & 8.411 & 10.703 \\
    \hline
    \multirow{4}{*}{\begin{tabular}[c]{@{}c@{}}Max diagonal\\ displacement\end{tabular}}   & R²  &\textbf{0.9716} & \textbf{0.9495}& 0.9071 & 0.4688   \\
    & MAE &\textbf{5.377}$\mathbf{\times 10^{-3}}$ & \textbf{9.122}$\mathbf{\times 10^{-3}}$& 5.581$\times 10^{-2}$ & 4.688$\times 10^{-1}$ \\
    & MSE &\textbf{9.348}$\mathbf{\times 10^{-5}}$ & \textbf{1.674}$\mathbf{\times 10^{-4}}$& 9.572$\times 10^{-3}$ & 1.459$\times 10^{-1}$ \\
    & MAPE (\%) & \textbf{2.824} &\textbf{4.883}  & 5.356 & 12.841 \\
    \hline
    \multirow{4}{*}{\begin{tabular}[c]{@{}c@{}}Max torsional\\ displacment\end{tabular}}   & R²  & \textbf{0.9704}&\textbf{0.9419}& 0.9223 & 0.7762  \\
    & MAE &\textbf{3.104}$\mathbf{\times 10^{-3}}$ & \textbf{4.955}$\mathbf{\times 10^{-3}}$& 3.136$\times 10^{-2}$ & 5.401$\times 10^{-2}$ \\
    & MSE &\textbf{3.021}$\mathbf{\times 10^{-5}}$ & \textbf{5.378}$\mathbf{\times 10^{-5}}$& 2.969$\times 10^{-3}$ & 6.589$\times 10^{-3}$ \\
    & MAPE (\%) & \textbf{2.259} &\textbf{3.862} & 12.794 & 23.667 \\
    \hline
    \end{tabular}
    \end{adjustbox}
\end{table*}

Synthetic data generated through augmentation techniques significantly enhance the surrogate models' performance. These data increase the model's generalization ability and prediction accuracy, demonstrating the value of using an augmented dataset. The key findings indicate that synthetic data improve the model's ability to generalize to new designs and the accuracy of performance predictions across a diverse design space.

\subsection{Validating Model Performance}
We utilized a test set uniformly sampled from both the latent and performance spaces to validate the performance of the constructed surrogate models. This test set includes representative samples from various design and performance categories, ensuring a comprehensive evaluation of the model's capabilities across different design scenarios. The evaluation uses the uniformly sampled test set to measure the efficiency of the surrogate models in predicting structural performance metrics.

The uniformly sampled test set, drawn from both the design and performance spaces, encompasses various design scenarios. This carefully selected set of unseen data allows for an accurate measurement of the model's generalization capabilities. The results from this evaluation reveal that the surrogate models developed using the DeepJEB dataset exhibit high predictive accuracy and reliability across various design and performance conditions. Furthermore, the analysis demonstrates that these models effectively capture the complex relationships between geometric features and structural performance metrics. The performance metrics, including mean absolute error (MAE), mean squared error (MSE), MAPE, and adjusted R² scores, indicate that the models provide precise predictions consistent with the expected performance trends. Furthermore, using the deep ensemble methodology with NLL loss enhances the models' robustness by quantifying prediction uncertainty, which is crucial for assessing the reliability of the models in practical applications.

This case study validates the effectiveness of the DeepJEB dataset and underscores the benefits of using synthetic data to improve model performance and reliability. The DeepJEB dataset provides a robust foundation for developing high-performance surrogate models. The ability of this dataset to generate diverse and high-quality data ensures that the models trained on this dataset can generalize well to new and unseen designs, enhancing their applicability in real-world engineering scenarios. This case study highlights the transformative potential of synthetic data in improving the accuracy and reliability of data-driven models in structural performance prediction.

\section{Conclusion}
\label{Conclusion}

This research addresses the limitations of traditional datasets in capturing complex geometries and performance metrics in structural engineering. By integrating advanced data generation techniques and rigorous validation processes, we developed the DeepJEB dataset, which significantly enhances the capabilities of surrogate models. This advancement demonstrates that high-quality synthetic datasets can effectively mitigate the constraints of existing data-driven models, enabling more accurate and efficient engineering designs.

Our primary experiment involved the creation and validation of the DeepJEB dataset, which comprises 2138 samples, is approximately 5.6 times larger than the SimJEB dataset. The DeepJEB dataset includes additional data not present in SimJEB, such as simulation results for second-order tetrahedral elements, signed von Mises stress data, modal analysis outcomes, and multi-view images. These enhancements provide a more comprehensive resource for various engineering applications.

In comparative evaluations, surrogate models trained on the DeepJEB dataset demonstrated superior performance over those trained on the SimJEB dataset. For example, in predicting maximum vertical displacement, the DeepJEB models achieved an improvement in R² by approximately 22.8\% on the test set, reducing the MAPE by over 77.3\% compared to SimJEB. Similar performance gains were observed across other engineering labels, such as horizontal, diagonal, and torsional displacements, indicating the dataset’s robustness and reliability in diverse design scenarios.

The detailed node-level field data in the DeepJEB dataset makes it particularly well-suited for advanced modeling techniques like GNN and high-dimensional convolutional networks. These models can leverage the intricate interactions and connectivity of each node to provide more precise and detailed predictions, enhancing the accuracy and applicability of surrogate models in structural engineering. Moreover, developing models for node value prediction (field prediction) opens up new avenues for understanding structural behavior, potentially leading to significant breakthroughs in the accuracy and reliability of engineering simulations.

In the future, updates to the dataset could include dynamic time series data obtained through the integration of crash simulations or multi-physics simulations. This would provide an even more comprehensive dataset that captures the complex, real-world behaviors of engineering systems, further enhancing the applicability and robustness of surrogate models.

In conclusion, the creation and validation of the DeepJEB dataset have significantly contributed to the field of mechanical engineering. Our research offers valuable insights and tools for enhancing data-driven engineering models by addressing the limitations of existing datasets and offering a robust framework for future dataset creation. This work not only improves the accuracy and reliability of surrogate models but also opens up new possibilities for their application in various engineering domains, driving forward innovation and progress in the field. As AI and data-driven approaches continue to evolve, datasets like DeepJEB will be crucial in pushing the boundaries of engineering research and practice, driving innovation and progress in the field.

\section{Licensing, attributions and access}
\label{Licensing}

The DeepJEB dataset is a derivative work incorporating synthesized CAD models based on a subset of ``The Simulated Jet Engine Bracket Dataset (SimJEB)'', licensed under the Open Data Commons Attribution License. The full license document can be accessed at reference \cite{odcby10}. Unlike the original SimJEB dataset, the DeepJEB dataset includes several modifications and enhancements, such as generating additional shapes through shape synthesis, including more comprehensive engineering information via additional simulation metrics, and providing multi-view images. These enhancements enable more versatile applications of the 3D bracket data in various engineering contexts. The DeepJEB dataset is also licensed under the Open Data Commons Attribution License.

The DeepJEB dataset is publicly accessible on Google Drive without any restrictions. This dataset can be accessed at the following URL: \href{https://www.narnia.ai/dataset}{https://www.narnia.ai/dataset}. Moreover, this repository provides all necessary files in standard formats, including simulation results (.csv), B-Rep CAD files (.step), second-order tetrahedral mesh files (.vtk), multi-view images (.png), tessellated surface mesh files (.stl), hierarchical data format files combining information on tetrahedral meshes and simulation results (.h5), train–test split index files (.json), and FEM solver input files (.fem).





\bibliographystyle{asmems4}
\bibliography{asme2e}

\end{document}